\documentclass[aps,prb,twocolumn,showpacs,amsmath,amssymb,superscriptaddress]{revtex4-1}
\usepackage{graphicx}
\usepackage{xcolor}
\usepackage{color}
 \usepackage[bookmarks=false]{hyperref}
\hypersetup{
    colorlinks = true,
   linkcolor ={black}
}
\begin{document}

\title{Route Towards Dirac and Weyl Antiferromagnetic Spintronics}

\author{L. \v{S}mejkal}
\affiliation{Institut f\"ur Physik, Johannes Gutenberg Universit\"at Mainz, D-55099 Mainz, Germany}
\affiliation{Institute of Physics, Academy of Sciences of the Czech Republic, Cukrovarnick\'{a} 10, 162 53 Praha 6 Czech Republic}
\affiliation{Faculty of Mathematics and Physics, Charles University in Prague,
Ke Karlovu 3, 121 16 Prague 2, Czech Republic}
\author{T. Jungwirth}
\affiliation{Institute of Physics, Academy of Sciences of the Czech Republic, Cukrovarnick\'{a} 10, 162 53 Praha 6 Czech Republic}
\affiliation{School of Physics and Astronomy, University of Nottingham, Nottingham NG7 2RD, United Kingdom}
\author{J. Sinova}
\affiliation{Institut f\"ur Physik, Johannes Gutenberg Universit\"at Mainz, D-55099 Mainz, Germany}
\affiliation{Institute of Physics, Academy of Sciences of the Czech Republic, Cukrovarnick\'{a} 10, 162 53 Praha 6 Czech Republic}


\maketitle

\twocolumngrid

\textbf{
Topological quantum matter and spintronics research have been developed to a large extent independently. 
In this Review we discuss a new role that the  antiferromagnetic order has taken in combining topological matter and spintronics.
This occurs due to the complex microscopic symmetries present in antiferromagnets that allow, e.g., for topological relativistic quasiparticles and the newly discovered N\'{e}el spin-orbit torques to coexist.  We first introduce the concepts of topological semimetals and spin-orbitronics. Secondly, we explain the antiferromagnetic symmetries on a minimal Dirac semimetal model and the guiding role of \textit{ab initio} calculations in  predictions of examples of Dirac, and Weyl antiferromagnets: SrMnBi$_{\text{2}}$, CuMnAs, and Mn$_{\text{3}}$Ge. 
Lastly, we illustrate the interplay of Dirac quasiparticles, topology and antiferromagnetism  on: (i) the experimentally observed quantum Hall effect in EuMnBi$_{\text{2}}$,  (ii) the large anomalous Hall effect in Mn$_{\text{3}}$Ge, and (iii) the theoretically predicted topological metal-insulator transition in CuMnAs. 
}

\begin{figure}[thb]%
  \includegraphics[width=.48\textwidth]{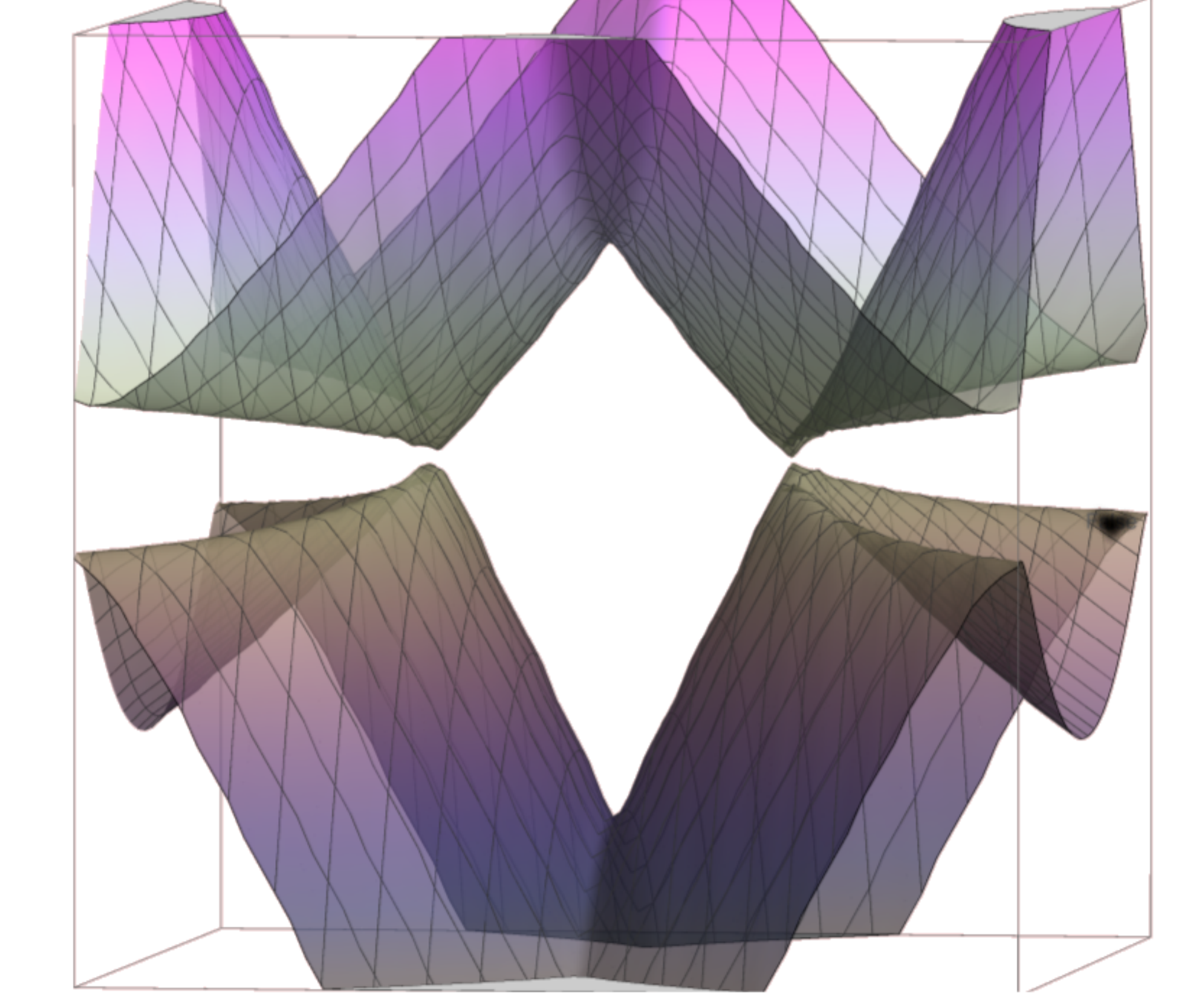}
\caption{Dirac fermions at the Fermi level of the Dirac semimetal antiferromagnet calculated from the fist-principles. Reorientation of the N\'{e}el vector drives the topological metal-insulator transition.}

\label{Fig_front}

\end{figure}

\onecolumngrid

\vspace{20pt}

\twocolumngrid

\tableofcontents

\begin{figure*}[t]%
  \includegraphics*[width=.63\textwidth]{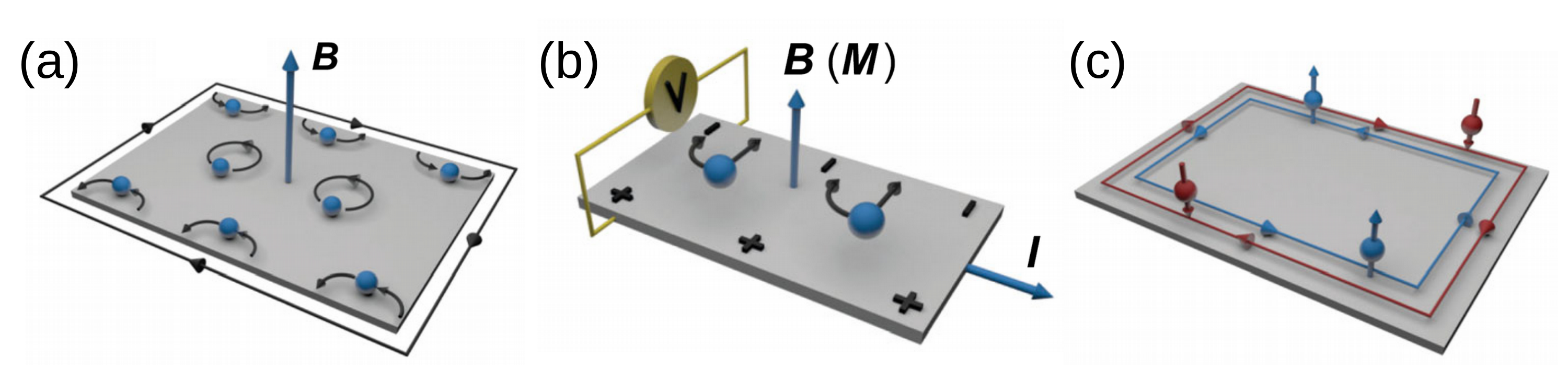}
\caption{(a) Quantum Hall effect and edge modes. (b)  Intrinsic anomalous Hall effect arising due to the time-reversal symmetry breaking by the magnetic order. (c) Quantum spin-Hall effect, or a 2D TI, exhibits two chiral edge states for the spin up and down. Reprinted 
from \cite{Sinova2015}.}
\label{Fig_Hall}
\end{figure*}

\section{Introduction: Topology meets spin}
In 1905 special relativity revolutionized physics. 
Almost one century later, the observation of relativistic-like effects and relativistic quasiparticles in solids has created a new revolution in condensed matter physics. 
In 2004 the discoveries of graphene \cite{Novoselov2004}  and topological insulators (TIs) \cite{Kane2005}
reignited the exploration of Dirac quasiparticles in solids and the search for novel topological states of matter \cite{Moore2010a}. 
Remarkably, also in 2004, the observation of the spin Hall effect  \cite{Kato2004d,Wunderlich2004} reinvigorated spintronics by shifting its focus from  non-relativistic effects towards effects originating  from  spin-orbit coupling (SOC). 
A culminating example of this is the recent phenomenon of the spin-orbit torque (SOT), that can be used to efficiently manipulate magnets \cite{Chernyshov2009,Miron2011b,Liu2012,Ciccarelli2016}.

Although relativistic (Dirac and Weyl) quasiparticles and TIs matured together, the  spintronics effects originating from SOC were developed to a large extent independently of them. 
However topology and certain spintronics effects are, at least on the theoretical level, very entangled. 
The intrinsic contributions to the spin Hall family of effects can be described in terms of topological properties of the wave functions  \cite{Sinova2015}. 
Recent works have began to combine directly spintronics, topological and Dirac materials   \cite{Hesjedal2016,Fert2013,Fan2016b}. 
Strong magnetic fields have been used to tune Dirac quasiparticle currents
in bulk layered antiferromagnets (AFs) at low temperatures \cite{Masuda2016}. 
In another example,  TIs have been used to enhance the efficiency of SOTs in TI/ferromagnetically (FM) doped TI heterostructures \cite{Fan2014a,Manchon2014b,Fan2016,Fan2016b}.

Unfortunately, most of the topological effects are still constrained to  low dimensionalities, high external magnetic fields, and very low temperatures  \cite{Moore2010a,Fan2016b}. In this review we show that antiferromagnetism combined with Dirac quasiparticles might become the missing ingredient on the route towards fully exploiting the potential of  topological spintronics. This promising perspective is provided by (i) the recent demonstration of the manipulation of AFs by electrical currents  \cite{Wadley2016}, (ii) the complex AF symmetries compatible with spintronics effects and nontrivial topologies that allow for Dirac and Weyl quasiparticles  \cite{Smejkal2016,Yang2017}, and (iii)  the externally magnetic invisibility in conjunction with the antiferromagnetic  order persisting above the room temperature offering novel functionalities \cite{Jungwirth2016}. We will show that the coupling between the AF order and relativistic quasiparticles not only leads to novel emergent effects but also pushes their limits. For instance, the recently proposed topological anisotropic magnetoresistance (AMR) in CuMnAs AF \cite{Smejkal2016} can be thought as a limiting case of the crystalline AMR.
The present review aims at providing the state-of-the art insight into the most recent developments from the point-of-view of theory and experiment. As such, it is not meant to be an exhaustive review of this fast developing field. 


\subsection{Topology and Hall effects}
The phases of matter can be classified by the Landau symmetry breaking paradigm \cite{Moore2010a}. For instance,  any crystalline phase is distinguished by the rotational, translational or other symmetry breaking by the crystal, or antiferromagnetic phase is determined by breaking the rotational symmetry of spins by the staggered order orientation. In the 1980s it was discovered that topology adds additional labeling to the phases \cite{Moore2010a}. For instance, two insulators in the same crystallographic group can be either topologically trivial or nontrivial \cite{Hasan2010}. Two band structures are topologically equivalent if they can be transformed into each other under continuous deformations. Topology considers spatial relationships between objects that survive these continuous transformations. In contrast, symmetry, as an invariance of the system under a given transformation, is described by group theory. 

Topology entered solid state physics in the seminal works on 2D phase transitions \cite{Kosterlitz1972,Kosterlitz1973}, quantum Hall effect (QHE) \cite{Klitzing1980}, and quantization of transport, and 
\cite{Thouless1982,Thouless1983}. The intrinsic Hall conductivity can be calculated according to the linear-response theory \cite{Sinova2015}: 
\begin{equation}
\sigma_{xy}=\frac{e^{2}}{\hbar }\sum_{n}\int_{\text{BZ}}\frac{d^{3}k}{(2\pi)^{3}}f_{n}(\textbf{k})\Omega_{n}(\textbf{k}),
\label{Eq_Kubo}
\end{equation}
where the summation is over  all occupied bands, $f_{n}(\textbf{k})$ is the Fermi-Dirac distribution function, and 
\begin{equation}
\Omega_{n}(\textbf{k})=2\text{Im}\sum_{m\neq n}\frac{\left\langle  \partial_{\textbf{k}_{x}} H(\textbf{k}) \right\rangle_{nm}\left\langle  \partial_{\textbf{k}_{y}} H(\textbf{k}) \right\rangle_{mn}}{\left(E_{n}(\textbf{k})-E_{m}(\textbf{k})\right)^{2}}. 
\label{Eq_BerryLRT}
\end{equation}
Here $\left\langle  \partial_{\textbf{k}_{x}} H(\textbf{k}) \right\rangle_{nm} =\left\langle u_{n}(\textbf{k})\vert \partial_{\textbf{k}_{x}} H(\textbf{k}) \vert u_{m}(\textbf{k}) \right\rangle$, where $H(\textbf{k})$ is the Hamiltonian with the corresponding eigenenergies $E_{n}$, and eigenvectors $\vert u_{n}(\textbf{k}) \rangle$, where $n$ is the eigenstate quantum number and \textbf{k} is the wave vector. The explicit link between topology of the wave function and the Hall conductivity can be made by recognizing that $\Omega(\textbf{k})\equiv \Omega^{z}(\textbf{k})$ is the $z$-component of the Berry curvature \cite{Nagaosa2010}:
\begin{equation}
\boldsymbol\Omega(\textbf{k})= \nabla_{\textbf{k}} \times i\left\langle u(\textbf{k}) \vert \nabla_{\textbf{k}} \vert u(\textbf{k}) \right\rangle,
\label{Eq_Berry}
\end{equation}
where we have dropped the band index for brevity. The Berry curvature transforms under spatial inversion symmetry $\mathcal{P}$, and time-reversal symmetry $\mathcal{T}$ as:
\begin{align}
\mathcal{P}&:  \boldsymbol\Omega(-\textbf{k})=\boldsymbol\Omega(\textbf{k}), \\
\mathcal{T}&: \boldsymbol\Omega(-\textbf{k})=-\boldsymbol\Omega(\textbf{k}).
\label{Eq_Berry_symmetry}
\end{align}
To obtain a nonzero transverse charge conductivity, it is necessary to break the time reversal symmetry. Otherwise the Berry curvature is an odd function of \textbf{k}  and, according to Eq. \eqref{Eq_Berry_symmetry} and \eqref{Eq_Kubo}, the Hall conductivity vanishes. For example, an external magnetic field does the job. When, additionally, a strong magnetic field is exerted on a two-dimensional (2D) electron gas, the Landau levels arise and in the insulating state for the Fermi level between fully occupied quantized levels the Hall conductivity \eqref{Eq_BerryLRT} becomes quantized:
\begin{equation}
\sigma_{xy}=\frac{e^{2}}{\hbar}\int_{BZ}\frac{d^{2}k}{(2\pi)^{2}}\Omega^{z}(\textbf{k})=\frac{e^{2}}{h}C.
\label{Eq_Chern}
\end{equation}
Here $C$ is the integer Chern number quantifying the underlying nontrivial topology of the edge state wave functions within the Landau level gap, provided the bands are well separated. This is the QHE, schematically illustrated in Fig.~\ref{Fig_Hall}(a).  
 
However, the nonzero Berry curvature \eqref{Eq_Berry} and Hall conductivity \eqref{Eq_Kubo} can be provided also by time reversal symmetry breaking due to an internal magnetic order. This is the anomalous Hall effect (AHE), illustrated in Fig.~\ref{Fig_Hall}(b), originating from the SOC in the time-reversal symmetry broken solid. The time-reversal symmetry breaking is explicitly given by the magnetization $M_{z}$ in ferromagnets (FMs), where it was empirically  established  for the Hall resistivity that \cite{Nagaosa2010}:
\begin{equation}
\rho_{xy}=R_{0}H_{z}+R_{S}M_{z}.
\label{hall_exp}
\end{equation} 
Here the first term corresponds to the ordinary Lorentz force related Hall effect, while the second term is the AHE contribution. Electron spin rotation during the adiabatic movement in the \textbf{k}-space leads due to the SOC to the nonzero Berry phase accumulation. The AHE in real solids has often important extrinsic contributions which we leave here aside \cite{Nagaosa2010}.
 
\begin{figure*}[bt]%
  \includegraphics*[width=.7\textwidth]{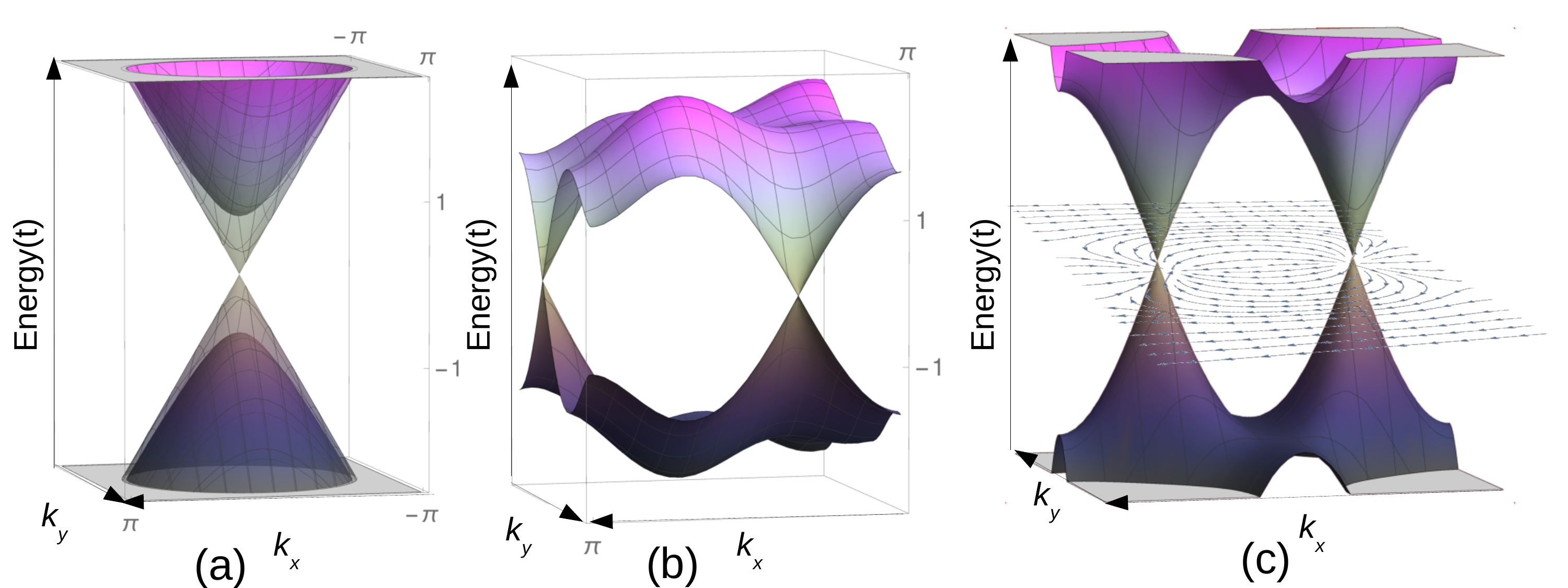}
\caption{\textbf{Schematic evolution of Dirac materials research.} (a) Bulk band inverted dispersion of the TI with nontrivial metallic surface states. The energy is in units of the hopping parameter $t$ (see Sec. 2.2). (b) Bulk Dirac quasiparticles protected by the nonsymmorphic crystalline symmetry reside at the BZ edges. (c) Bulk Weyl points act as magnetic mono-poles connected by the Berry curvature stream-lines.}
\label{Fig_Weyl}
\end{figure*}

In the late 1980s, Haldane showed the possibility of the QHE without Landau levels in the presence of complex hopping magnitudes on the honeycomb lattice \cite{Haldane1988}. This was the first conceptual step towards the quantum spin-Hall effect (QSHE) \cite{Kane2005}. The QSHE, or the 2D TI, are two copies of the QHE for the spin-up and spin-down \cite{Hasan2010}. The discovery of graphene and 2D surface states of 3D TIs  brought Dirac quasiparticles explicitly into the game \cite{Hsieh2008}. In contrast to the original QHE, edge states of 3D TIs are topological states protected by the time  reversal symmetry, and can be thus in principle gapped. The effective surface Hamiltonian acquires the form of the Dirac equation \cite{Zhang2009a}:
\begin{align}
\mathcal{H}(\textbf{k})=\hbar v_{F}\left( \sigma_{x}k_{y}-\sigma_{y}k_{x} \right),
\label{surfaceDirac}
\end{align}
where $\sigma$ are Pauli matrices and $v_{F}$ is the Fermi velocity. A surface Dirac cone is shown in Fig.~\ref{Fig_Weyl}(a). Topological transport exhibits robust dissipation-less edge states which are protected from the back-scattering and thus have been considered as ideal platforms for applications in spintronics, as we describe in Sec.1.3  \cite{Hasan2010,Fan2016b}.  Unfortunately, the effects are typically constrained to  high magnetic fields, low temperatures, or reduced dimensionality, and it is challenging to find proper material candidates. For instance, the unintentional bulk doping in TIs, and the challenge to make TIs compatible with magnetism, hinders their full potential for spintronics \cite{Sun2016a,Fan2016b}.
\subsection{Dirac and Weyl semimetals}
Nontrivial topologies and relativistic quasiparticles can be also associated with the bulk degeneracies in the band structure \cite{Vafek2013,Burkov2016}. In the  prototypical Dirac quasiparticle system -- graphene -- Dirac points acquire a nonzero mass, i.e., the band crossing is avoided, due to the SOC \cite{Kane2005}.
Unavoided band crossings were investigated from the very early days of quantum theory  \cite{Herring1937}, and the existence of the limiting phases of matter between insulators and metals was consider already in 1970s \cite{Abrikosov11971}. The recent rejuvenation of the interest in the bulk degeneracies due to the experimental discoveries of relativistic semimetals  \cite{Neupane2013,Liu2014e,Xu2015b,Lv2015} was made possible only by the identification of  suitable material candidates based on the state-of-the-art first-principle electronic structure theory   \cite{Wang2012f,Wang2013g,Huang2015,Weng2015}. 

Electrons in conventional crystals form typically Schr\"{o}dinger bands. 
However, when the bands cross accidentally close to the Fermi level in crystals with specific symmetries, they might create a relativistic semimetal phase. Fermi states in relativistic semimetals are dominated by the emergent relativistic  quasiparticles similarly to graphene. Adopting the high energy physics terminology, the relativistic particles come in three flavors: Weyl, Dirac and Majorana fermions \cite{Vafek2013}. In solids, the classification of the effective electronic quasiparticles is much richer due to the fact these quasiparticles do not have to obey the relativistic Lorentz symmetry, while they might be constrained by additional crystalline symmetries, not present in high energy physics  \cite{Wieder2016,Bradlyn2016}.  Here we focus on Dirac and Weyl quasiparticles in AFs, as is depicted in Fig.~\ref{Fig_Weyl}(b-c), and \ref{MDirac}(b-e). 

Dirac quasiparticles are allowed in systems with doubly-degenerate bands  \cite{bernevig2013topological}. Within the single particle picture, the eigenvalue $E_{n\sigma}(\textbf{k})$ and the eigenvector $\psi_{n\sigma}(\textbf{k})$ of the Hamiltonian of the solid $H_{0}$, are labeled by the quantum number $n$, spin $\sigma$, and the crystal momentum $\textbf{k}$ in the first Brillouin zone (BZ).  Double-band degeneracy is realized in systems invariant under the combined  spatial inversion $\mathcal{P}$, and time-reversal $\mathcal{T}$ symmetries.  The $\mathcal{T}$ symmetry acts as: $ E_{n,\uparrow}(\textbf{k})=E_{n,\downarrow}(\textbf{-k}) $, while $\mathcal{P}$ acts as: $ E_{n,\sigma}(\textbf{k})=E_{n,\sigma}(\textbf{-k})$ giving rise to $E_{n,\uparrow}(\textbf{k})=E_{n,\downarrow}(\textbf{k})$ over the whole BZ. Let us then consider the $\mathcal{PT}$ invariant solid with two double degenerate bands well separated from the rest of the band structure. The corresponding effective single particle Hamiltonian restricts to the four-band Hamiltonian: $H_{0} \rightarrow \mathcal{H}_{\text{eff}}=\sum_{\textbf{k}n,m,\sigma,\sigma' }
 \psi^{\dagger}_{ n \sigma}(\textbf{k}) \mathcal{H}(\textbf{k}) \psi_{ m\sigma'}(\textbf{k}),$ where $\psi^{\dagger}_{ n\sigma}(\textbf{k})$ creates a particle in a state $\vert u_{n\sigma}(\textbf{k}) \rangle$, and the matrix elements are given by $\mathcal{H}_{mn\sigma\sigma'}(\textbf{k})=\left\langle u_{n\sigma}(\textbf{k}) \vert H_{0} \vert u_{m\sigma'}(\textbf{k})  \right\rangle$.  When we choose  $\mathcal{P}= \tau_{x}$, $\mathcal{T}=-i\sigma_{y}\mathcal{C}$ ($\mathcal{C}$ being complex conjugation), and we restrict the Hamiltonian by $\left[\mathcal{H},\mathcal{P}\mathcal{T}\right]=0$, we obtain:
\begin{equation}
\mathcal{H}(\textbf{k})=\sum_{k=0}^{5}A_{j}(\textbf{k})\Gamma_{j},
\label{genDirac}
\end{equation}
where the Dirac matrices $\Gamma_{j}=\left\lbrace 1, \tau_{x}, \tau_{y}, \sigma_{x}\tau_{z}, \sigma_{y}\tau_{z}, \sigma_{z}\tau_{z} \right\rbrace$ and $A_{j}(\textbf{k})$ are functions of crystal momentum. 
The energy spectrum is then given by:
\begin{equation}
E_{\pm}(\textbf{k})=A_{0}(\textbf{k})\pm \sqrt{\sum_{k=1}^{5}A_{j}^{2}(\textbf{k})},
\label{dispersion}
\end{equation}

To ensure the stable accidental band-crossing (ABC), the expression under the square root must vanish \cite{Vafek2013,Yang2014a}. In general, it is not possible to tune simultaneously five functions  $A_{j}(\textbf{k})$ to zero by varying just three components of the crystal momentum $\textbf{k}$. We can reduce the number of free functions to three by additional crystalline symmetries, which can further reduce the number of $\Gamma$ matrices in the Hamiltonian \eqref{genDirac}. 
By abandoning the $\tau_{y}$ term due to the additional crystalline symmetry, neglecting $\sigma_{0}$ terms that tilt and shift the bands, assuming isotropic Fermi velocities $v_{F}$ and keeping the  $\tau_{x}$ term constant we obtain in the vicinity of the ABC the effective Dirac Hamiltonian:  
\begin{align}
\mathcal{H}(\textbf{k}) = 
\left(\begin{matrix} 
 \hbar v_{F}(\textbf{k}-\textbf{k}_{0})\cdot \boldsymbol\sigma & m \\
m &  -\hbar v_{F}(\textbf{k}-\textbf{k}_{0})\cdot \boldsymbol\sigma
\end{matrix}\right).
\label{MDirac}
 \end{align} 
Recently it was demonstrated that the protection mechanism can by provided by nonsymmorphic \cite{Young2012,Fang2015,Young2015,Schoop2016} or rotational symmetries \cite{Wang2012f,Wang2013g,Yang2014a}. The nonsymmorphic symmetry is a combination of a point group symmetry with a nontrivial translation \cite{bradley2010mathematical}. The crystalline symmetry $\mathcal{S}$ prevents some terms in the Hamiltonian by  $\left[\mathcal{H},\mathcal{S}\right]=0$. On the \textbf{k}-subspace invariant under $\mathcal{S}$, the bands can be labeled eventually by the symmetry eigenvalues preventing the hybridization, as we show in Sec. 2. Thus, {symmetry-protected} Dirac quasiparticles can be found typically at the rotational axes \cite{Yang2014a} or the BZ edges \cite{Young2012,Young2015}, as we illustrate in Fig.~\ref{Fig_Weyl}(b). In this model 2D Dirac semimetal, the Dirac crossings at the $X$ points are protected by nonsymmorphic symmetries \cite{Young2015}.

Magnetic or non-centrosymmetric crystals have non-degenerate bands by violating $\mathcal{PT}$ symmetry by breaking
$\mathcal{T}$\cite{Wan2011,Wang2016c}, or $\mathcal{P}$\cite{Weng2015,Huang2015}, or both symmetries \cite{Zyuzin2012,Chang2016b}. The low energy physics around the ABC comprising two non-degenerate bands can be approximated by a two-band Hamiltonian:
\begin{equation}
\mathcal{H}(\textbf{k})=\sum_{i=0}^{3}A_{i}(\textbf{k})\sigma_{i}\, .
\end{equation} 
Similarly as in the four-band case, omitting the $A_{0}$ term and expanding the Hamiltonian  in the vicinity of the ABC gives:
\begin{equation}
\mathcal{H}(\textbf{k})=\hbar v_{\text{F}}(\textbf{k}-\textbf{k}_{0})\cdot \boldsymbol\sigma,
\label{Weyl}
\end{equation}
where we have additionally imposed an isotropic Fermi velocity $v_{\text{F}}$ for the sake of recovering the explicit form of the Weyl equation known from high-energy physics. The topology is now reflected in the fact that the 3D Weyl Hamiltonian \eqref{Weyl} uses its complete bases - all three Pauli matrices. Thus, any sufficiently small perturbation will just shift, but not gap, the ABC in the BZ \cite{Burkov2016}.  The Weyl fermions come always in pairs with opposite chirality, as can be seen by breaking the $\mathcal{PT}$ symmetry in Eq.\eqref{MDirac} and as was generically proven for fermions by the \textit{no-go theorem} by Nielsen and Ninomiya  \cite{Nielsen1981,Witten2015}. Weyl points can be gapped only by annihilation with another Weyl point of the opposite chirality. Weyl fermions, the fundamental building blocks of the standard model, were never observed in high energy physics. In 2015, they were observed in a non-centrosymmetric crystal TaAs  \cite{Xu2015b,Lv2015}. The Weyl points are typically found inside the BZ, as can be seen on the band structure of the  $\mathcal{PT}$ symmetry breaking Weyl semimetal model  in  Fig.~\ref{Fig_Weyl}(c)  \cite{Yang2011b,Vafek2013}, but can reside also at the edges \cite{Chang2016}. The nontrivial topology of the electronic wave function can be quantified by the topological index of the band-crossing similarly to Eq. \eqref{Eq_Chern}. The Berry curvature stream lines for the time-reversal braking Weyl semimetal model \cite{Yang2011b,Vafek2013} are depicted in the  Fig.~\ref{Fig_Weyl}(c). The Chern number is calculated, in contrast to  Eq.\eqref{Eq_Chern}, as the integral over the Berry curvature \eqref{Eq_Berry} on the small sphere around the Weyl point\cite{Gresch2016}: 
\begin{equation}
C=\frac{1}{2\pi}\int_{S}dS\Omega(\textbf{k})=\pm 1\, .
\end{equation}
The Chern number value $\pm1$ is realized for the linear Weyl crossing and corresponds to the opposite sign of the chirality of the two Weyl points in Fig.~\ref{Fig_Weyl}(c), with chirality defined as the projection of the spin to the momentum axis. The Berry curvature  in the vicinity of the Weyl point is approximated as:
\begin{equation}
\boldsymbol\Omega(\textbf{k})=\pm\frac{\textbf{k}}{2k^{3}}\,,
\end{equation}
which explicitly shows that the Weyl points act as a source and a drain of the Berry curvature. As we see in Fig.~\ref{Fig_Weyl}(c),  the Berry curvature stream lines can be used to track the position of the Weyl points since the Weyl points act as effective monopoles of the Berry curvature. The streamlines of the Berry curvature connect the two Weyl points in a similar form as the magnetic field lines connect the north and the south pole of a magnet. Thus, a single Weyl point can be thought of as an analog of the magnetic monopole.  The Berry flux between the Weyl points in the time-reversal breaking Weyl semimetal gives rise to a non-quantized anomalous Hall conductivity, $\sigma_{xy}=-\frac{w}{\pi}\frac{e^{2}}{h},$ where $w$ is the separation of the Weyl points in the BZ \cite{Sun2016a}.
 
Topological Weyl and Dirac semimetals are attractive due to  the possible robust symmetry and topology protected electronic states, topological surface states - Fermi arcs \cite{Wan2011,Vafek2013,Xu2015,Xu2015b,Huang2015,Yang2015d}, presumably high mobilities \cite{Neupane2013,Shekhar2015}, or exotic magneto-transport phenomena such as the negative longitudinal magnetoresistance  \cite{Ali2014,Arnold2015,Liang2014,Shekhar2015} and other manifestations of the chiral anomaly \cite{Hirschberger2016,Jia2016,Zyuzin2012a}.  
The most investigated Dirac semimetals are nonmagnetic: Na$_{\text{3}}$Bi \cite{Liu2014e}, and Cd$_{\text{3}}$As$_{\text{2}}$  \cite{Neupane2013}, or ZrSiS  \cite{Schoop2016}. The Weyl semimetal state was firstly observed in the TaAs family as mentioned above \cite{Xu2015b,Lv2015}. 
\begin{center}
\begin{figure*}[tb]%
\centering
  \includegraphics*[width=.764\textwidth]{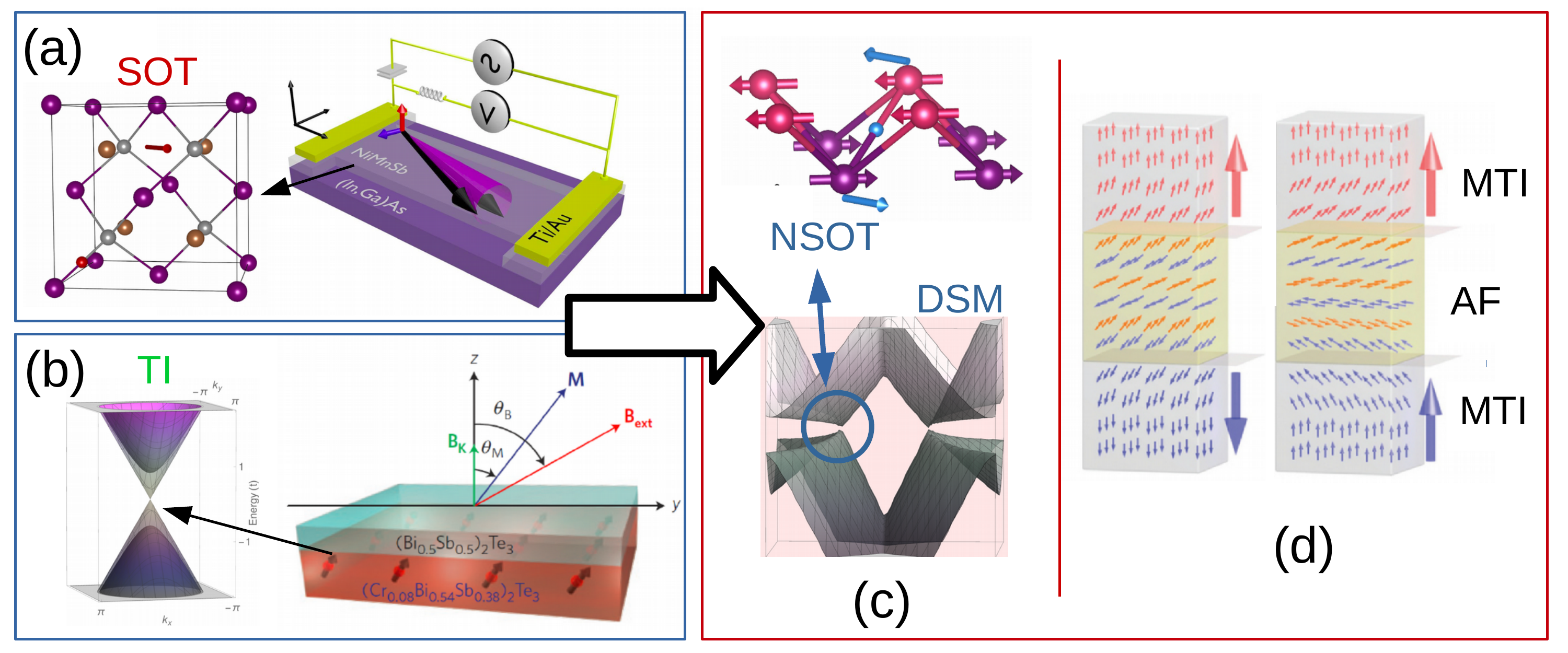}
  \caption[]{%
  \textbf{Schematic evolution of the spintronics research.}
   (a) Dirac approach to spintronics, e.g. spin-orbitronics, is rooted in the relativistic spin-orbit coupling. A typical example is the SOT due to the non-equilibrium spin polarization induced in a non-centrosymmetric crystals, such as NiMnSb (left panel). (b) Alternatively,  in spintronics based on TIs, the perfect spin-momentum locking of the surface states can torque the adjacent magnetically doped TI (MTI).  The next level of interplay of topology and AF can be realized via two approaches:  (i) in heterostructures, e.g., (d) the proximity coupling between MTI and AF, and (ii)  in AFs with specific microscopic symmetries, as the AF crystal in (c), there is an interplay between Dirac quasiparticles and the NSOT.
Fig.(b) and (d) are reprinted 
from \cite{Fan2014a,Hesjedal2016}.
    }
    \label{topospin}
\end{figure*}
\end{center}
\subsection{Spintronics and antiferromagnets} 
We first recall recent conceptual developments in spintronics  on selected prototypical devices and principles.
Traditionally the interaction between magnetization and conduction electron spin was modeled by assuming the s-d type of exchange interaction \cite{Jungwirth2014}.  
Prominent effects of this Mott approach \cite{Jungwirth2014} to spintronics include the giant magnetoresistance (GMR), and the spin-transfer torque (STT), which were explored in the FM/spacer/FM spin valve structures.
The GMR refers to changes in the resistivity induced by rotating the moment in the free FM layer, which can be used to read its state.  The GMR can be explained in terms of the Mott two current model of transport in an exchange split band FM \cite{Jungwirth2014}. The switching of the magnetization in the free layer leads to the different scattering rates for the spin-up and spin-down channels, and consequently resistivity changes. 
The STT can be used for writing the magnetic information. In the STT, the fixed layer is used to spin-polarize the electric current. The spin polarized current then exerts a torque on the free magnetic layer \cite{Ralph2008}.

The discoveries of the spin Hall effect (SHE) and TIs shifted the focus of spintronics partly towards relativistic effects. 
Novel approaches to spintronics include concepts based on the interchange of spin and momentum due to the SOC in heavy metals ('Dirac principle' in  \cite{Jungwirth2014}), and concepts employing low dissipation Dirac quasiparticle surface states of TIs \cite{Fan2016b}, as we explain in Fig.~\ref{topospin}. 
The archetypal effects of the magnetic spin-orbitronics (Fig.~\ref{topospin}(a)) are the AMR and the SOT. The AMR and the SOT, in contrast, to the GMR and the STT, rely on the relativistic SOC. Due to the SOC, electrons feel different scattering rates for the magnetization oriented parallel or perpendicular to the electrical current direction, leading to the AMR \cite{Jungwirth2014}. 
The non-centrosymmetric magnet subjected to an electric current generates a non-equilibrium spin-orbit field, and thus the SOT, which in turn can reorient its magnetization. Recently this has been demonstrated even at room temperature  \cite{Ciccarelli2016}.
The spin-orbitronics path of spintronics research also led recently to the emergence of spintronics based on AFs \cite{MacDonald2011,Jungwirth2016,Baltz2016}.  The applicability of  relativistic physics to AF based spintronics was demonstrated in the seminal works on AF AMR \cite{Park2011b,wang2014c,Marti2014,Wadley2016,Jungwirth2016}. 
The role of relativistic effects is even more pronounced in the AF spintronics since the GMR  in AFs remains elusive \cite{Jungwirth2016}, while comparable AMR signals to FMs were observed in AF semiconductors \cite{Park2011b,wang2014c}, metals \cite{Marti2014} and recently also semimetals \cite{Wadley2016}. 
The lack of practical means for the
 manipulation of AF moments and their microscopic complexity 
left for decades AFs as primarily passive elements in spintronics devices providing magnetic pinning of the reference FM layer. The breakthrough  was the demonstration of the electric  current manipulation of the AF order
 in a CuMnAs semimetal by the N\'{e}el SOT (NSOT) \cite{Zelezny2014,Wadley2016}.  

In  topological based spintronics (Fig.~\ref{topospin}(b)) the TI is interfaced with a magnet \cite{Fan2016b}. The combination of the perfect spin-momentum locking and the strong SOC of TIs can lead to large spin accumulation or spin current under the applied electric current \cite{Fan2016b}. The Dirac quasiparticle spin on the surface of the TI is locked to be perpendicular to the particle momentum according to Eq.\eqref{surfaceDirac}.  When the TI is subjected to the lateral electric field, the charge current with a spin polarization is generated at the surface due to the spin-momentum locking in combination with a Fermi surface shift from the Dirac point
 \cite{Fan2016b}. For example, in Fig.~\ref{topospin}(b), a negative charge current in the [100] direction would generate a spin-polarization in the [0$\overline{1}$0] direction.  
 This spin polarization can be then used to exert a torque on the magnetization, as was recently demonstrated in a TI/magnetically-doped TI heterostructure \cite{Fan2014a}. Here it was shown that by reversing the direction of the lateral current, the magnetization of the Cr-doped TI can be switched under the assistance of the external magnetic field of 0.6 T \cite{Fan2014a}. Very efficient switching with a critical current $8.9.10^{4}$A/cm$^{2}$ -- three order of magnitudes lower than in heavy-metal/FM bilayers -- was achieved \cite{Fan2016b}. However, the  enhanced efficiency is constrained to low temperatures.  The main obstacle is that the magnetic order in TIs  is stable only at very low temperatures.
The subfield is looking for suitable high temperature material candidates which would be possible to efficiently couple with TIs. To this end the suitability of the AF order was demonstrated recently \cite{He2016}. An efficient proximity effect at room temperature was demonstrated in a CrSb superlattice AF sandwiched between TI  heterostructures, as depicted in Fig.\ref{topospin}(d)  \cite{He2016,Hesjedal2016}. 
 
The recently proposed novel approach to spintronics based on topological AFs combines the benefits of both topological states and the SOC driven spintronics. Here the antiferromagnetic order would outperform FMs. In Ref.~ \cite{Smejkal2016} it was theoretically suggested that the topological effects, Dirac quasiparticles, and SOTs can join forces in the specific class of AFs due to their unique symmetries, which are not present in FMs.  In Fig.~\ref{topospin}(c) we depict the concept based on a topological AF semimetal. The working principle is the control of the Fermi surface topology by manipulating the AF order parameter by the electric current, i.e., by the NSOT. The reading of the magnetic state can be achieved due to the predicted effect inherent to topological AFs, topological AMR (TopoAMR) \cite{Smejkal2016}. It originates from controlling the symmetry protection of Dirac points by the N\'{e}el vector direction.  In topological AFs, spintronics effects can be pushed to their limits. For example, the presence of relativistic quasiparticles can enhance the strength and efficiency of the SOT \cite{Ghosh2016,Hanke2017a}. Vice versa, the SOT manipulation of the N\'{e}el order parameter can be used to tune the masses of Dirac fermions, which can lead, in principle, to a topological metal-insulator transition (TopoMIT), and the aforementioned TopoAMR. One possible material realization was suggested recently in the orthorhombic phase of the CuMnAs AF semimetal  \cite{Smejkal2016}, as we will explain in more detail in Sec. 4.3.
\begin{figure*}[htb]%
  \includegraphics*[width=.57\textwidth]{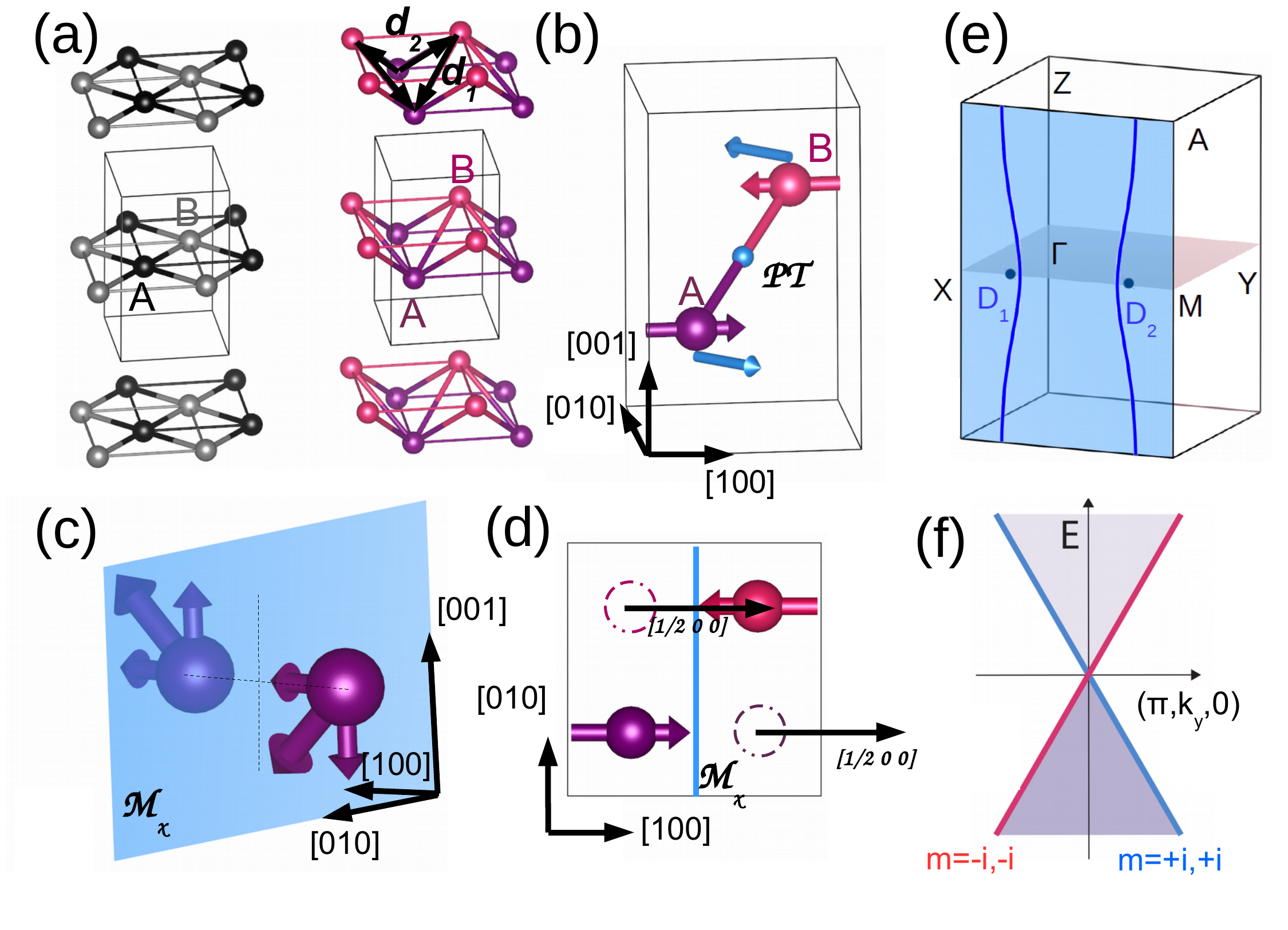}%
  \caption[]{%
  \textbf{Minimal model for Dirac antiferromagnetic semimetal.} (a) The crinkled lattice of the quasi-2D AF Dirac semimetal is built from the square AF layers by shifting the $A$ and $B$ atoms in the opposite direction along the [001] axis. (b)
  Unit cell of the AF Dirac semimetal model with the effective antiferromagnetic time reversal symmetry $\mathcal{T}_{\text{eff}}=\mathcal{PT}$  and the staggered non-equilibrium spin-polarization (blue arrows). (c) Mirror picture of the axial vector, e.g., magnetization, spin, Berry curvature. (d) Glide-mirror plane $\left\lbrace M_{x}\vert \left( \frac{1}{2}, 0, 0 \right) \right\rbrace$ of the model AF. (e) Corresponding invariant subspace in the BZ (blue) and the position of the two Dirac points. (f) The $\mathcal{M}_{x}$ symmetry assignments around the Dirac point $D_1$ in Fig.\ref{fig_tb}(b). 
}
    \label{Fig_sym}
\end{figure*}

\section{Theory of Dirac and Weyl antiferromagnets}
\subsection{Antiferromagnetic symmetry}
AFs owe many of their unique properties to their external magnetic invisibility in combination with internal magnetic long range order  \cite{Jungwirth2016}.
The external magnetic invisibility is given by the defining feature of AFs, the zero net magnetic moment. 
The AF symmetries can also lead to the existence of some effects (Dirac quasiparticles, NSOT), but at the same time they can make other effects vanish (the AHE in simple collinear AFs, the NSOT in centrosymmetric AFs). 
AFs can yield more physical phenomena than FMs when the effective AF symmetries rescue the topological phases, e.g., AF TIs \cite{Mong2010,Fang2013}, or AF topological  Dirac semimetals  \cite{Smejkal2016,Tang2016}. 
AF TIs rely on the effective time reversal symmetry $T_{\frac{1}{2}}\mathcal{T}$, which combines the time reversal symmetry $\mathcal{T}$ with a half-magnetic unit cell translation $T_{\frac{1}{2}}$  \cite{Mong2010}. This is
not possible in FMs.
The Dirac semimetal, as we explained in Sec. 1.2, can be realized only in systems with doubly degenerate bands. In any magnetic crystal, the time-reversal symmetry $\mathcal{T}$ is explicitly broken. 
In FMs there is no symmetry operation which can rescue the double-band degeneracy once the time-reversal symmetry is broken globally by the net magnetization. 
Remarkably, in AFs where both $\mathcal{T}$ and spatial inversion symmetry $\mathcal{P}$ are broken, but their combination $\mathcal{P}\mathcal{T}$ is preserved, the double band degeneracy over the whole BZ is reinstated   \cite{Herring1966,Chen2014,Tang2016,Smejkal2016}. Consequently, $\mathcal{P}\mathcal{T}$ AFs might host  Dirac quasiparticles. 
The Dirac semimetal state can be stabilized in bands associated with eigenvalues of crystalline symmetry operators at the corresponding BZ invariant subspaces \cite{Young2012,Wang2012f,Tang2016,Smejkal2016}.  In the recently proposed AFs, the symmetry protection of these relativistic quasiparticles is due to the nonsymmorphic symmetries \cite{Tang2016,Smejkal2016}. 
In general,  magnetism hugely enlarges the playground for studying the fundamental relationship between symmetry and  topology in solids due to the necessity to consider 1651 magnetic space groups instead of the 230 nonmagnetic ones  \cite{bradley2010mathematical}.
The general classification of the magnetic and nonsymmorphic symmetries is beyond the scope of this brief review. 
Instead, we illustrate the physics governed by the AF symmetries on a minimal tight binding model introduced recently  \cite{Smejkal2016,Young2015,Yang2016b}.

\subsection{Minimal Dirac semimetal antiferromagnet}
In a recent work \cite{Smejkal2016} it was shown that the interplay between spintronics and topology can be explained on a minimal Dirac AF semimetal model which, additionally,  can serve as a parent phase for the massive Dirac fermions or the magnetically induced Weyl semimetal. 
The generic lattice band Hamiltonian with the symmetry of Eq.~\eqref{MDirac} can be formulated by considering two AF sublattices with one orbital and a spin per atom on the square lattice, as in Fig.~\ref{Fig_sym}(a):
\begin{equation}
\mathcal{H}=\sum_{\left\langle i,j \right\rangle,\left\langle \left\langle i,j \right\rangle\right\rangle}t_{ij}\hat{c}_{i}^{\dagger}\hat{c}_{j}+\sum_{i}J_{i}\hat{c}_{i}^{\dagger}\textbf{n}\cdot\boldsymbol\sigma \hat{c}_{i},
\label{tbmodelAF}
\end{equation}
where $c_{i}$ is the annihilation operator, $t_{ij}$ is the hopping amplitude between nearest $\left\langle i,j \right\rangle$ and next nearest neighbor $\left\langle \left\langle i,j \right\rangle\right\rangle$ atoms, $J_{i}$ is the AF exchange, and \textbf{n} is the unit N\'{e}el vector. 
\begin{figure*}[htb]
  \begin{center}
  \includegraphics*[width=1.\textwidth]{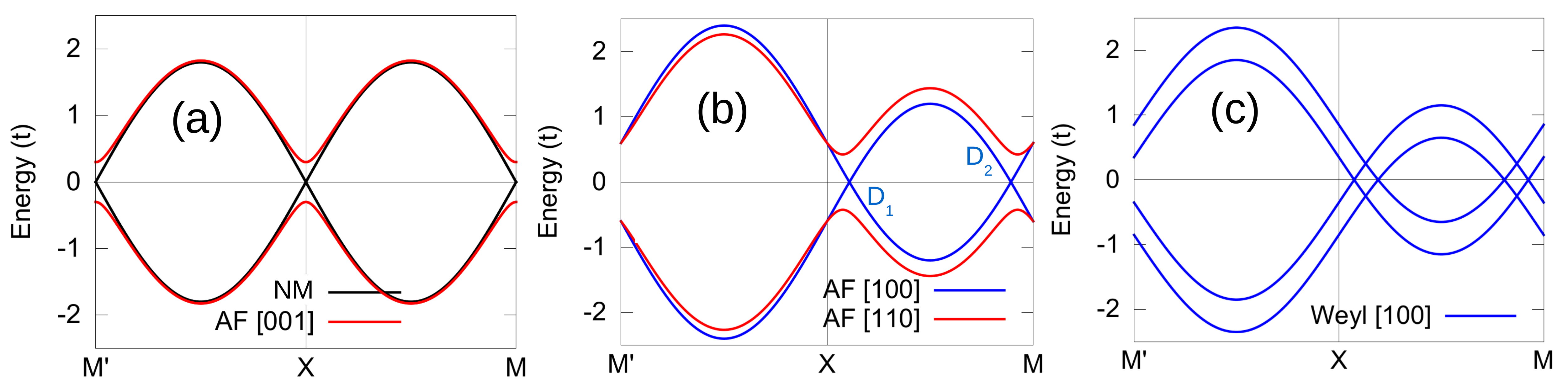}%
  \caption[]{
  \textbf{Band structure of the minimal Dirac semimetal AF model. }   (a) Dispersion of the nonmagnetic (black line) model. AF order along [001] introduces masses to the Dirac fermions (red line). (b) Easy-axis AF along [100] preserves mirror off-centered nonsymmorphic symmetry which protects Dirac points along the $M-X$ line. (c) An external magnetic field applied along the [100] splits the Dirac points into four Weyl points, leading to the magnetically induced Weyl semimetal.    
}
    \label{fig_tb}
      \end{center}
\end{figure*} 
The coexistence of Dirac quasiparticles, nontrivial topologies, and the NSOT can arise by deforming the above square layered lattice in the left panel of Fig.~\ref{Fig_sym}(a) to the nonsymmorphic crystal in the right panel. The  second neighbor, or Kane-Mele \cite{Kane2005,Young2015}, $\textbf{k}$-dependent SOC is introduced by moving the $A$ and $B$ atoms in the opposite direction along the [001] axis,
\begin{equation}
\mathcal{H}_{\text{SOC}}(r) =i\sum_{\left\langle \left\langle i,j \right\rangle\right\rangle,\left\langle k \right\rangle} \lambda_{ij}\hat{c}_{i}^{\dagger}\left(\textbf{d}_{ik}^{1}\times \textbf{d}^{2}_{kj} \right) \cdot \boldsymbol\sigma \hat{c}_{j},
\label{KM}
\end{equation}
where $\lambda_{ij}$ is the SOC strength, and $d^{1,2}_{ik}$ are bonds to the nearest atom interconnecting the next nearest neighbor atoms as illustrated in Fig.~\ref{Fig_sym}(a). The SOC term in Eq.~\eqref{KM} is a double Rashba Hamiltonian since the $A$ ($B$) atom has the nearest inter-layer $B$ ($A$) atom. In this sense, the term is staggered analogously as the AF Zeeman term - the second term in Eq. \eqref{tbmodelAF}. This double staggering results into band-inversion and band-touching at specific sections of the BZ.
The \textbf{k}-space Hamiltonian is obtained by the Fourier transformation and has the structure of  Eq.\eqref{genDirac}, where the coefficients $A_{j}(\textbf{k})$ have now direct physical meaning  \cite{Smejkal2016}. We will further discuss the quasi-2D case, where the inter-layer hopping is neglected assuming the much larger inter-layer than intra-layer distances.  The Hamiltonian reduces to $\mathcal{H}(\textbf{k})=\sum_{j=1,3,4,5}A_{j}(\textbf{k})\Gamma_{j}$, with the nearest neighbor hopping term $A_{1}(\textbf{k}) = -2t\cos \frac{k_{x}}{2} \cos \frac{k_{y}}{2}$ (crystal momentum \textbf{k} is in dimensionless units), and the combined SOC/exchange terms $A_{3}(\textbf{k})=J_{x}-\lambda \sin k_{y}$, $A_{4}(\textbf{k})=J_{y}+\lambda \sin k_{x}$, and $A_{5}(\textbf{k})=J_{z}$. By neglecting the next nearest neighbor hopping, the spectrum has an additional particle-hole symmetry  \cite{Young2015}.
The energy spectrum is given by Eq.\eqref{dispersion}.  

The relevant aspects of the band touchings can be minimalistically explained at the $M'-X-M$ line in the BZ, where we have chosen $t=1\,\text{eV}$, $J=0.6t$, and a large SOC parameter $\lambda=1.5t$. In  Fig.~\ref{fig_tb} (a) we plot the energy dispersion for a nonmagnet ($J=0$), depicted by the black line. The spectrum shows Dirac points at $X$ and $M$ protected by multiple symmetries of the nonmagnetic lattice in the right panel of Fig. \ref{MDirac}(a) \cite{Young2015}. 
The 3D band structure of the distorted version of this model is plotted in  Fig.~\ref{Fig_Weyl}~(b). For the sake of brevity, we plot here the band structure of the model with Dirac points only at  $X$. The Dirac point at $M$  are gapped by adding the term $\Delta_{1}\sin\frac{k_{x}}{2}\sin\frac{k_{y}}{2}$\cite{Young2015} to the hopping part $A_{1}(\textbf{k})$ which breaks rotational screw symmetries $C_{2x/y}$ protecting it. When the AF coupling (N\'{e}el vector) is switch-on along the [001] axis in the undistorted model, the Dirac points preserve their \textbf{k}-space location but they acquire a mass, as illustrated in Fig.~\ref{fig_tb}(a) by the solid red line.   

For the N\'{e}el vector along the [100]-direction, we have Dirac points in 2D, as is depicted in the Fig.~\ref{fig_tb}(b). By considering the inter-layer hopping, the 3D model is recovered \cite{Smejkal2016} and  the Dirac node lines appear in the $k_{x}=\pi$ plane (see schematics in Fig.~\ref{Fig_sym}(e)). The Dirac points in 2D and Dirac node lines in 3D are protected by the nonsymmorphic glide mirror plane symmetry\cite{Yang2016b,Smejkal2016}
$
\mathcal{G}_{x} =\left\lbrace M_{x} \vert \left(\frac{1}{2},0, 0 \right) \right\rbrace, 
$
which combines a nontrivial translation by $\left(\frac{1}{2}, 0, 0\right)$ with a mirror plane reflection $\mathcal{M}_{x}$, as depicted in Fig.~\ref{Fig_sym}(d). The crystal momentum transforms under $\mathcal{G}_{x}$ (note that in the reciprocal space only the point group part of  $\mathcal{G}_{x}$, namely $\mathcal{M}_{x}$ matters) as: 
$
\mathcal{G}_{x}\left(k_{x},k_{y},k_{z}\right)=\left(-k_{x},k_{y},k_{z}\right),
$
making the $k_{x}=0,\pi$ BZ sub-spaces invariant under $\mathcal{G}_{x}$,
The $\mathcal{M}_{x}$ symmetry is represented as, $\mathcal{M}_{x}=i\sigma_{x}\tau_{z}$, with the two crystal-momentum independent eigenvalues $m_{\pm}=\pm i$. The nonsymmorphicity of $\mathcal{G}_{x}$ is determined by the different center of symmetry for the $\mathcal{G}_{x}$ and $\mathcal{PT}$ symmetries, since
$
\mathcal{G}_{x}\circ\mathcal{PT}=e^{ik_{x}}\mathcal{PT}\circ\mathcal{G}_{x}.
$
The 2D Hamiltonian in the vicinity of  the Dirac point (DP) $D_1$ in Fig.~\ref{fig_tb}(b)  at the crystal momentum  $\textbf{D}=\left(\pi,\arcsin \frac{J}{\lambda}\right)$ can be expanded  as:
\begin{align}
\begin{split}
\mathcal{H}_{\text{eff}}(\textbf{d}_{1}+\textbf{k})=-2t\cos \frac{d_{y}}{2}k_{x}\tau_{x}- \\ 
-2\lambda\left( \cos d_{y} k_{y}\sigma_{x}+ k_{x}\sigma_{y} \right)\tau_{z}.
\end{split}
\label{dispersionkp}
\end{align}
At the $\mathcal{G}_{x}$ invariant subspace ($X-M$ line) the effective Hamiltonian in Eq.~\eqref{dispersionkp} has one degenerate eigenvalue $E_{+1,2}(k_{y})=2\lambda \cos d_{y}k_{y}$ with the corresponding 
eigenvectors: 
$
\left\langle u_{\textbf{k},+1}\right.\vert=\frac{1}{\sqrt{2}}\left(0, 0, 1, 1\right)$,
$\left\langle u_{\textbf{k},+2}\right.\vert=\frac{1}{\sqrt{2}}\left(-1, 1, 0,0\right)
$
and a second degenerate eigenvalue $E_{-1,2}(k_{y})=-2\lambda \cos d_{y}k_{y}$ with the corresponding eigenvectors: 
$
\left\langle u_{\textbf{k},-1}\right.\vert=\frac{1}{\sqrt{2}}\left( 0,0,-1,1 \right)$,
$\left\langle u_{\textbf{k},-2}\right.\vert=\frac{1}{\sqrt{2}}\left(1,1, 0,0 \right).$
The $\mathcal{G}_{x}$ 
symmetry shares eigenvectors with the Hamiltonian.
The $\mathcal{M}_{x}$ symmetry expectation values $m_{\pm}=\left\langle u_{\textbf{q},\alpha}\vert \mathcal{M}_{x} \vert u_{\textbf{q},\alpha}\right\rangle$ for the two states $E_{+1,2}$ are  $m_{+}=-i$, while for $E_{-1,2}$ are $m_{-}=+i$. Consequently  each double degenerate band belongs to  a different symmetry representation as depicted in Fig.~\ref{Fig_sym}(f). The eigenvalue correspondence to the bands protects the band crossing since it prevents hybridization \cite{Yang2014a}.

The interplay of Dirac fermions with the NSOT is  allowed due to the fact that the deformation of the sublattices has broken the inversion symmetry of the magnetic atom sites $A$ and $B$. The NSOT was 
theoretically predicted in Ref.~ \cite{Zelezny2014} and subsequently experimentally observed in current induced switching experiments in the tetragonal CuMnAs AF \cite{Wadley2016}. The microscopic physics of the NSOT is discussed in a recent article by \v{Z}elezn\'{y}  \cite{Zelezny2017}.
In Fig.~\ref{fig_tb}(c) we show the effect of the external magnetic field $\delta h=0.25t$ applied in the [100] direction. $\delta h$ splits the two Dirac points into four Weyl points at the Fermi energy.
  
With this minimal model we have shown here  that the AFs can generate topologically nontrivial states by themselves. This opens the prospect for spintronics devices with built-in topological states beyond those that combine TIs and AFs at interfaces. Next we review realistic topological AF (semi)metallic candidates from the perspective of  first-principle calculations.

\subsection{First-principle calculations of the electronic structure and transport effects}
Recent advances in the understanding of topological properties of condensed matter and spin-orbitronics effects were led, to a large extent, by the developments in density functional theory (DFT) calculations simulating the solids from a microscopic quantum-mechanical description \cite{Jones2015,bluegel2014computing}. 
The practical implementation of the theory requires a problem-suitable choice of the effective potential approximation (in the simplest form the local density approximation (LDA) or the generalized gradient approximation (GGA)) and of the wavefuction basis (typically plane waves or tight-binding).
Due to the variational formulation of the DFT, it is possible to determine  ground-state wave functions from a numerical self-consistent iteration process controlled, e.g., by the electronic density convergence. 
From the ground-state wave functions further quantities of interest are calculated.
To reveal nontrivial topologies of the surface states, calculations in the slab geometry are required and in the context of Dirac semimetals there is a need to determine the symmetries of the states comprising the band crossing. Plots of the projected local density of states, as the one in Fig.~\ref{mn3ge}(c), then allow to distinguish the topology and the character of the surface states and can help to understand the experimental data, e.g., angle resolved photo-emission spectroscopy or scanning tunneling microscopy (see for instance Refs. \cite{Yang2015d,Batabyal2016}). 

For spintronics applications, transport effects are simulated based on the linear response theory. In this formalism, the transport coefficient can be obtained from the response function of the ground-state wave functions, typically in the form similar to  Eq.~\eqref{Eq_BerryLRT} (see e.g. Refs. \cite{Gradhand2012,Freimuth2015,Zelezny2017}). 
In the presence of Dirac quasiparticles in the band structure, the computational costs increase due to the typically dominating contribution from the (un)avoided crossings. Additionally, in the realm of spintronics, it is desirable to simulate the transport under realistic conditions, e.g., at finite temperature and in imperfect crystals. Recent advances in this direction include the {\it ab initio} inclusion of finite temperature effects \cite{Liu2011c,Kodderitzsch2013}, or the multi-scale approach, where the {\it ab initio} results only parametrize the atomistic model \cite{Janson2014}. The possibility of simulating on  equal footing strong relativistic and correlation effects in topological magnetic semimetals  can be perceived as an exciting future direction in  computational physics. Dynamical mean field theory combined with the LDA is the established starting point \cite{Li2017}. All the results presented in the next section were obtained  within the standard GGA (+SOC) full potential linearized augmented plane-wave method with the Perdew-Burke-Ernzerhof parametrization \cite{Perdew1996}.

\section{Electronic structure of Dirac and Weyl antiferromagnetic candidates}
In this section we give a summary of three key AFs which have been shown to exhibit relativistic quasiparticles or are potential candidates. 
\subsection{Dirac antiferromagnets $X$MnBi$_{\text{2}}$}
\begin{figure}[htb]%
  \includegraphics*[width=.5\textwidth]{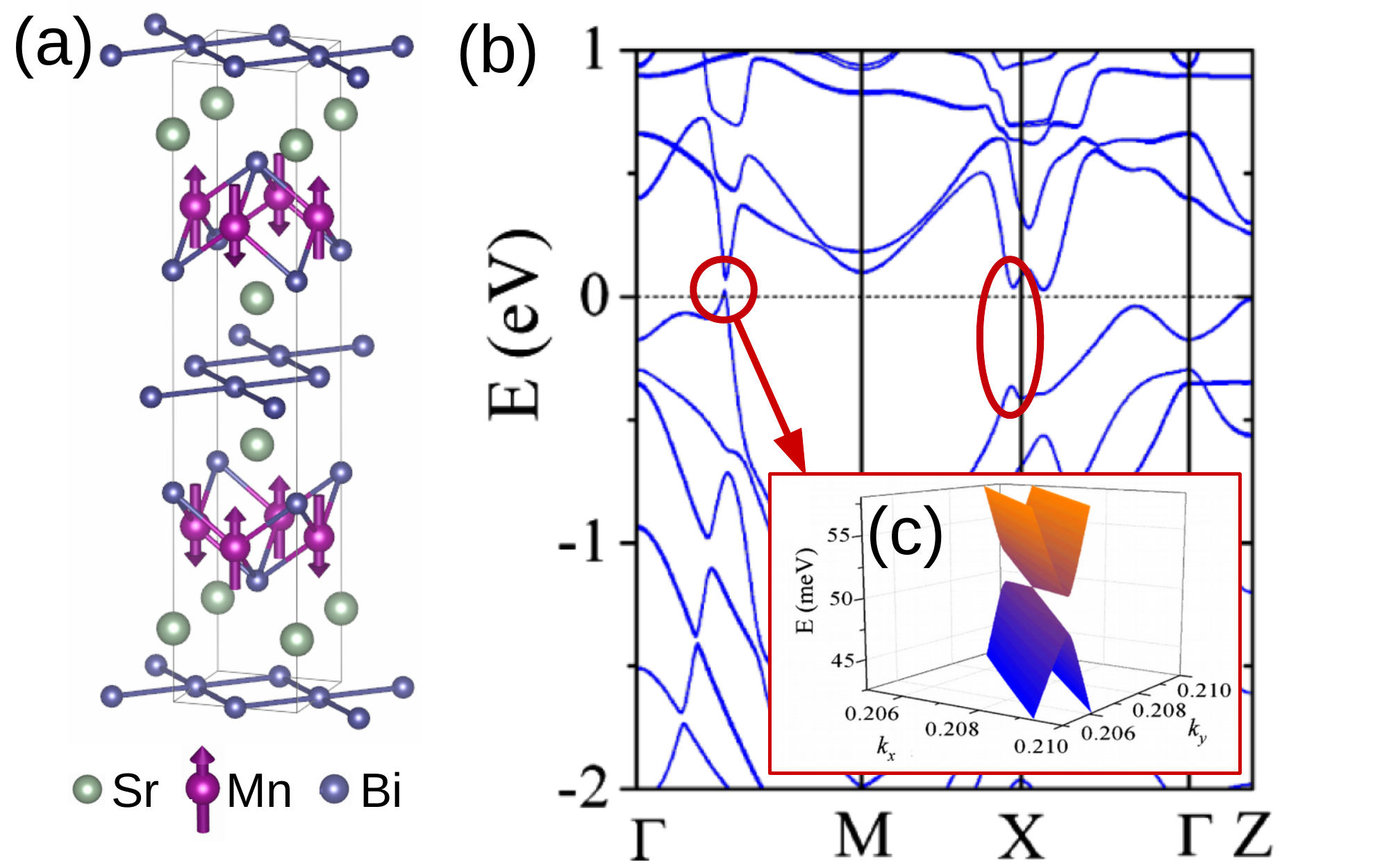}
  \caption[]{%
  \textbf{Dirac antiferromagnet SrMnBi$_{\text{2}}$.} (a) The crystal consist of  quasi-2D layers of Sr, Mn, and Bi. (b) Band structure calculated with SOC shows the massive Dirac fermions around Fermi level (in red ellipses).  (c) Detail of anisotropic Dirac fermion along $\Gamma - M$ calculated from SrBi layer subsystem without SOC. (b-c) adapted 
  from \cite{Park2011a}.    
    }
    \label{srmnbi2}
\end{figure}

Layred AFs from the 112 pnictides family, 
CaMnBi$_{\text{2}}$ and SrMnBi$_{\text{2}}$ \cite{Park2011,Lee2013b}, support the double-band degeneracy of the electronic bands. In Fig.~\ref{srmnbi2} we show the crystallographic structure of  SrMnBi$_{\text{2}}$ consisting of alternating Sr, Mn, and Bi layers. 

\begin{figure*}[bt]%
  \includegraphics*[width=.7\textwidth]{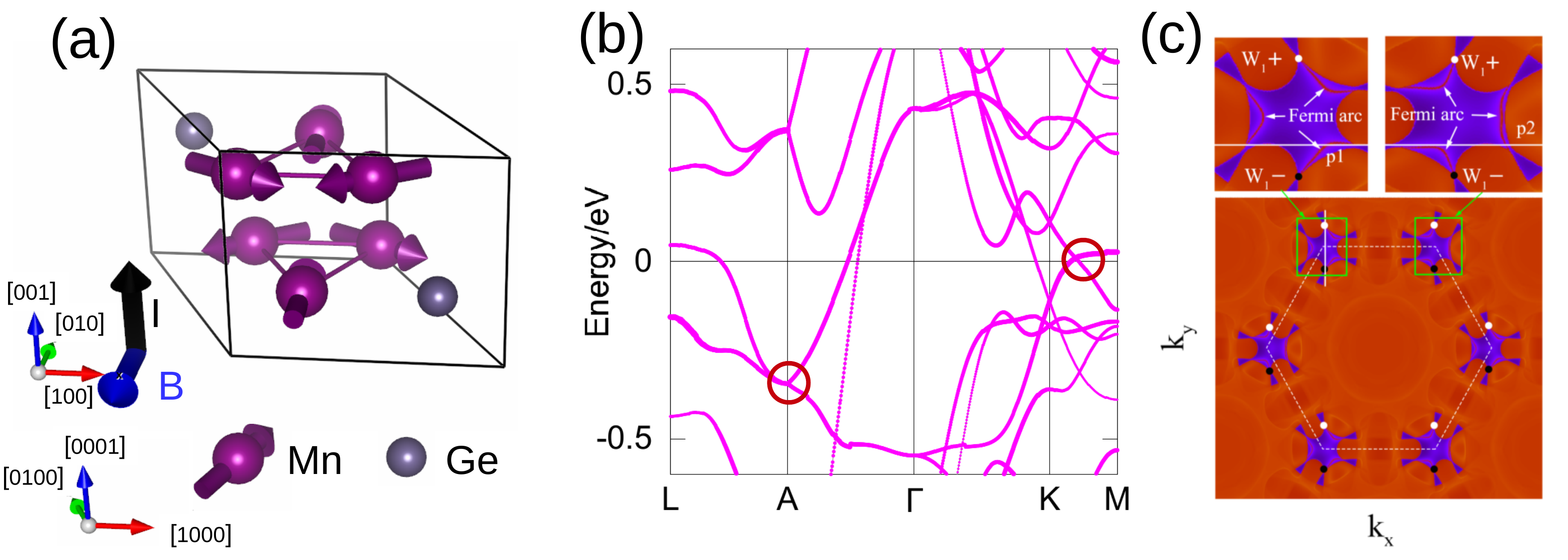}%
  \caption{\textbf{Electronic structure of the non-collinear Weyl antiferromagnet Mn$_{\text{3}}$Ge.} (a) Crystallographic structure of the non-collinear AF Mn$_{\text{3}}$Ge comprises from the kagome lattices stacked along the [001] axis. (b) Band structure of Mn$_{\text{3}}$Ge calculated from {\it ab initio}. (c) \textit{Ab initio} calculation of the topological Fermi arc surface states. Figure (c) adapted  
from  \cite{Yang2017}.}
    \label{mn3ge}
\end{figure*} 
The band structure calculated with the SOC (Fig.~\ref{srmnbi2}(b)) shows massive Dirac fermions \cite{Park2011a}, similar to the case of the minimal model with the N\'{e}el vector along the [001] direction.
The XMnBi$_{\text{2}}$ AF family thus represents the massive quasi-2D-Dirac fermion system. Most of the electronic properties are determined by the Bi square sheets, as can be seen from the orbital composition \cite{Park2011a,Lee2013b,Guo2014}. 

The quasi-2D character is reflected in the almost flat dispersion along the $\Gamma-Z$ line in the BZ (c.f. Fig.~\ref{srmnbi2}~(b)). Additionally, the AF exhibits a quasi-2D quantum transport, namely a highly anisotropic angular dependent magnetoresistance  \cite{Wang2011e}. Note that this very large magnetoresistance is given by the projection of the magnetic field on the Fermi surface and is distinct from the spontaneous AMR driven by the SOC in magnets, which is a representative effect of the Dirac spintronics. The different structure of the Dirac cones in SrMnBi$_{\text{2}}$ (Fig.~\ref{srmnbi2}(c)) and CaMnBi$_{\text{2}}$ was attributed to the different position of the Sr and Ca atom in combination with a different magnitude of the SOC \cite{Lee2013b,Guo2014,Feng2014}. Both AFs have the N\'{e}el temperature around room temperature. In contrast, the EuMnBi$_{\text{2}}$  has   also an AF coupled Eu sublattice at very low temperatures \cite{Masuda2016}. The control of the AF order at the Eu sublattice leads to the transport effects discussed in Sec 4.1. A more detailed review on the topic of pnictides (also regarding superconductivity) can be found in Ref.~\cite{Ray2016}.
\subsection{Weyl antiferromagnets Mn$_{\text{3}}X$} 
The first realistic Weyl semimetal candidate was predicted in the pyrochlore AF Y$_{\text{2}}$Ir$_{\text{2}}$O$_{\text{7}}$, based on DFT  \cite{Wan2011}. This AF Weyl semimetal has not been confirmed yet despite substantial experimental effort  \cite{Sushkov2015}. An alternative candidate with properties appealing to spintronics, namely strong AHE, was predicted recently in chiral non-collinear AFs Mn$_{\text{3}}$X (X=Ge,Sn)  \cite{Yang2017}.
In  Fig.~\ref{mn3ge}(a) we show the crystallographic structure built from stacked kagome planes along the [001] axis. 
The materials are known to have relatively weak magnetic anisotropy, reaching approximately $\sim$\,0.1\,meV for the Mn$_{\text{3}}$Sn  \cite{Sandratskii1996,Duan2015} and a net magnetic moment of 0.005\,$\mu_{\textbf{B}}$ per unit cell \cite{Tomiyoshi1982}. The predicted triangular magnetic structures in Mn$_{\text{3}}$Sn shown in  Fig.~\ref{mn3ge_exp}(b,c) \cite{Tomiyoshi1982,Sandratskii1996} were supported also by torque measurements of the magnetic anisotropy  \cite{Duan2015}. 
The crystal in its magnetic texture in Fig.~\ref{mn3ge_exp}(b) has a glide mirror plane $\mathcal{G}_{y}=\left\lbrace \mathcal{M}_{y} \vert \left( 0,0,\frac{1}{2} \right) \right\rbrace$, and two effective time reversal symmetries combining mirror symmetries $\mathcal{M}_{x}\mathcal{T}$ and $ \mathcal{M}_{z}\mathcal{T}$. Any of these three symmetries double the number of Weyl points leading to the multiplicity of 8 \cite{Yang2017}.
Band structure calculations in  Mn$_{\text{3}}$Ge reveal several Weyl points around the Fermi level together with other trivial states as illustrated in Fig.~\ref{mn3ge}(b-c). The small symmetry breaking due to the net moment  corrects slightly the position of the Weyl points related by the mirror symmetries in the ideal AF structure with  zero net moment  \cite{Yang2017}. The detailed position of the Weyl points was located by tracking the Berry curvature in the whole BZ as explained in Sec. 1.2. The hallmark of Weyl semimetal states, the nontrivial Fermi arc surface states, was predicted by first-principle calculations of the local density of states and are depicted in  Fig.~\ref{mn3ge}(c)  \cite{Yang2017}. 
The study published in this issue of the PSS reveals that in spite of the weak anisotropy the inverted chiral structure is relatively stable against thermal fluctuation and it is possible to influence the in-plane chiral AF magnetic structure by the spin-filtering effect \cite{Fujita2016}.
\begin{figure*}[hbt]%
  \includegraphics*[width=.9\textwidth]{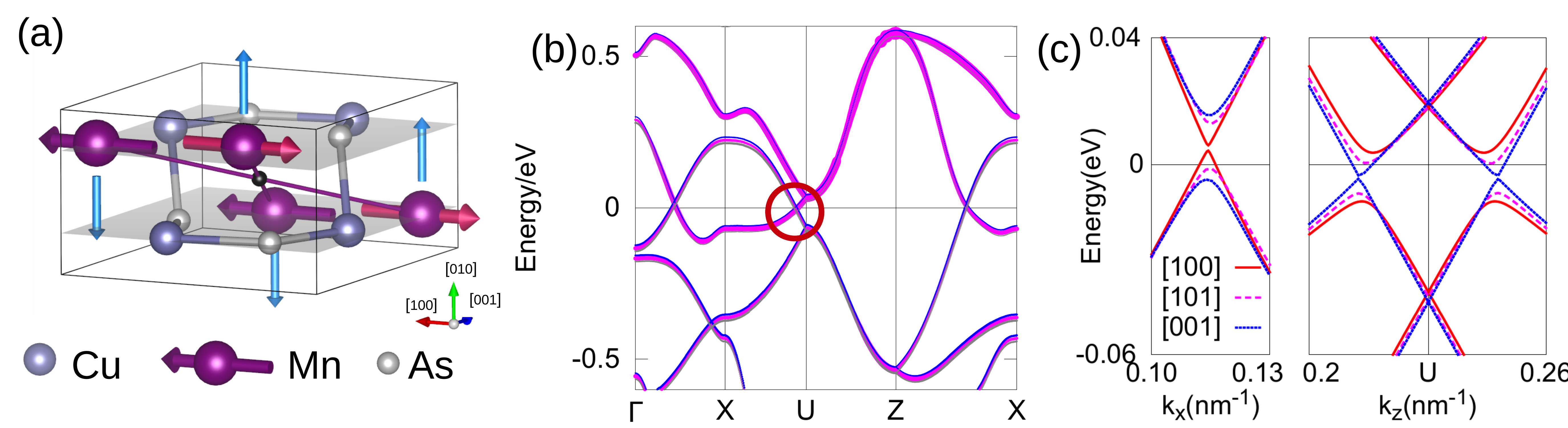}
\caption{\textbf{AF Dirac semimetal.} (a) Orthorhombic crystal structure of the CuMnAs consists of two $\mathcal{PT}$-Mn partners. (b) The atom-resolved band-structure shows the dominat contribution from Mn (magenta color) to the band-inversion at the Fermi level. (c) Topological metal-insulator transition in CuMnAs driven by N\'{e}el order reorientation. 3D variant of the Dirac semimetal for the \textbf{n}$\parallel$[001] is depicted in the figure in the abstract. 
}
\label{cumnas}
\end{figure*}
\subsection{Dirac semimetal antiferromagnets CuMn$X$} 
CuMnX (X=As,P) has been originally studied in its orthorhombic form  as a promising AF semiconductor candidate \cite{Maca2012}. Electronic structure calculations and transport measurements point towards a semimetallic phase \cite{Maca2012,Tang2016,Smejkal2016}. The tetragonal phase was used to experimentally discover the NSOT  \cite{Wadley2016}. Very recently, a symmetry protected Dirac semimetal state was predicted in the orthorhombic phase of CuMnX AFs \cite{Smejkal2016,Tang2016}. 
 
In CuMnAs the Dirac points can carry topological charges and are protected by the combined $\mathcal{PT}$ symmetry together with a certain nonsymmorphic symmetry, in analogy with the minimal model in Sec. 2.2. $\mathcal{PT}$ symmetry ensures a double band degeneracy over the whole BZ \cite{Herring1966,Chen2014,Tang2016,Smejkal2016}, while the nonsymmorphic symmetry prevents hybridization of the bands at the band-crossing  \cite{Tang2016,Smejkal2016}. The nonsymmorphic pattern in orthorhombic CuMnAs is slightly more involved than in the minimal model. Orthorhombic CuMnAs contains four Mn sublattices that are connected in pairs by the $\mathcal{PT}$ symmetry, as seen in  Fig.~\ref{cumnas}(a).
The atomic resolved band structure calculated without the SOC is depicted in  Fig.~\ref{cumnas}(b) and shows dominating Mn orbitals at the Fermi level. Three visible Dirac points at the Fermi level along the $\Gamma - X$, $X - U$, and $Z - X$ lines are part of the node line related to the glide mirror plane symmetry $\mathcal{G}_{y}=\left\lbrace \mathcal{M}_{y} \vert \left( 0,\frac{1}{2},0 \right) \right\rbrace$ \cite{Tang2016}. The protected Dirac semimetal realized for the N\'{e}el order along the [001] axis is shown in Fig.~\ref{cumnas}(c), and appears by gapping the nodal line except for the two Dirac points along the $U - X - U$ subspace. The Dirac points are protected by the nonsymmorphic screw axis  
$\mathcal{S}_{z}=\left\lbrace 2_{z} \vert \left( \frac{1}{2},0,\frac{1}{2} \right) \right\rbrace$  \cite{Tang2016,Smejkal2016} and are connected via nontrivial surface states  \cite{Tang2016}. The topological index of the band-crossing can be defined for the AF semimetal in analogy to the nonmagnetic Dirac semimetals \cite{Smejkal2016,Yang2016b}.

The topological Dirac semimetal in CuMnAs is appealing, since theoretically only a pair of Dirac points occurs at the Fermi level according to \textit{ab initio} calculations \cite{Tang2016,Smejkal2016}, thus offering an ideal model for a topological Dirac semimetal induced by band-inversion \cite{Bernevig2015,Wang2016c}. In the next Section we will review the recent prediction of merging spintronics with topology and the novel magnetotransport effects in CuMnAs. 

\section{Interplay between topology and antiferromagnetism}
Topological AFs can bring effects that cannot take place in either nonmagnets or FMs. 
We review here the interplay of the Dirac quasiparticles, the QHE, and antiferromagnetism in the ternary pnictides. We also review the intrinsic contribution from Weyl quasiparticles to the giant anomalous Hall effect in the non-collinear chiral AF Mn$_{\text{3}}$Ge, and novel  effects predicted for  CuMnAs. 

\subsection{Interplay of Dirac quasiparticles, antiferromagnetism and quantum Hall effects in ternary pnictides}
The interaction between Dirac quasiparticles and magnetism was demonstrated in ternary pnictides, although the Dirac quasiparticles and magnetism arise from different physical origins \cite{Guo2014}. 
Recently, enhancement of the exchange coupling between layers via Dirac carriers in  CaMnBi$_{\text{2}}$ and SrMnBi$_{\text{2}}$ was found with the help of Raman scattering \cite{Zhang2016g}. Also the magnetic field manipulation of the transport was demonstrated in the sister compound EuMnBi$_{\text{2}}$ \cite{Masuda2016}.
Magnetism can open an energy gap at the Dirac points in CaMnBi$_2$ and SrMnBi$_2$, which was attributed to a FM inter-layer coupling of Mn moments in CaMnBi$_2$ and to an AF coupling in SrMnBi$_2$ \cite{Guo2014}. The different behavior of the two compounds is due to the competing AF super-exchange and the FM double exchange mediated by the itinerant Bi electrons \cite{Guo2014}. 
The presence of the 40 meV band gap at the Dirac point along the $\Gamma - M$ line can lead potentially to a large contribution to the spin-Hall effect \cite{Park2011a} as was discussed for the similar paramagnetic situations in the iron-based superconductors \cite{Ray2016} and Weyl semimetal TaAs \cite{Sun2016a}. 

\begin{figure}[htb]%
\includegraphics*[width=\linewidth]{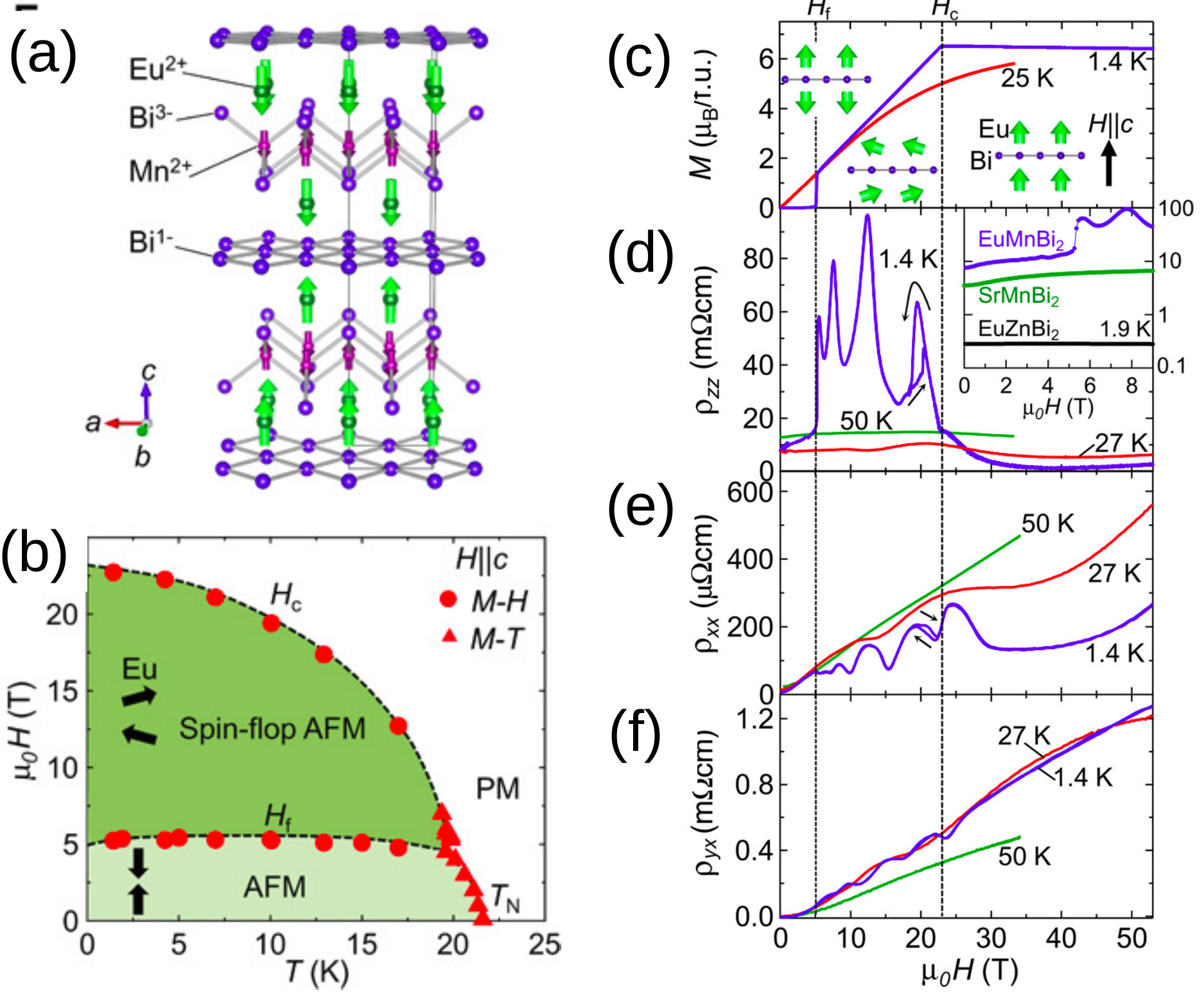}
\caption{\textbf{Quantum Hall effect in EuMnBi$_{\text{2}}$ manipulated by strong magnetic fields.} (a) Crystallographic and magnetic structure. (b) Phase diagram temperature-magnetic field. (c-f) The magnetization, the out-of-/in- plane, and the transversal resistivity show clearly quantum Hall effect for the finite window of the applied magnetic field. Masuda et al. \cite{Masuda2016} attributed the effect to the manipulation of the staggered order at the Eu site. Adapted from Ref.~ \cite{Masuda2016}}
\label{EuMnBi2}
\end{figure}

The Eu-based compounds behave in large magnetic fields differently to SrMnBi$_{\text{2}}$, due to the additional AF ordering on the Eu moments. The  suppression of the carrier density was attributed to the AF order of the Eu atoms  and demonstrates the  influence of magnetism on the Fermi surface \cite{Masuda2016}. We have shown in the preceding section the Dirac bands close to the Fermi level in  SrMnBi$_{\text{2}}$. Similarly, quasi-2D Dirac fermions are expected also in the EuMnBi$_{\text{2}}$. Dirac fermions presumably give rise to a large positive linear magnetoresistance, as can be seen in Fig.\ref{EuMnBi2}(e), and high mobilities up to 10 000 cm$^{\text{2}}$/Vs \cite{Masuda2016}. The influence of the magnetic field on the transport and magnetic properties of EuMnBi$_{\text{2}}$  is reproduced in Fig.~\ref{EuMnBi2}. In Fig.~\ref{EuMnBi2}(b) we show the phase diagram typical for the external magnetic field applied along the easy axis in an anisotropic AF. From the net magnetization measurement we see that above the spin-flop field, the Eu moments reorient perpendicular to the the applied field, while above the spin-flip field the moments order ferromagnetically. The AF ordering of Eu moments has substantial influence on the inter-layer transport, as can be seen from Fig.~\ref{EuMnBi2}(d-f). Furthermore, a half-integer QHE was reported in EuMnBi$_{\text{2}}$  controllable by the strength of an external magnetic field \cite{Masuda2016}. The QHE was attributed to the sufficient suppression of the [001]-axis conductivity and confinement of the massive Dirac fermions to the Bi-square quasi-2D layers by the spin-flop at the Eu sites \cite{Masuda2016}. Nevertheless, the detailed mechanism has not yet been identified.
\begin{figure*}[t]%
\begin{center}
\includegraphics*[width=0.9\linewidth]{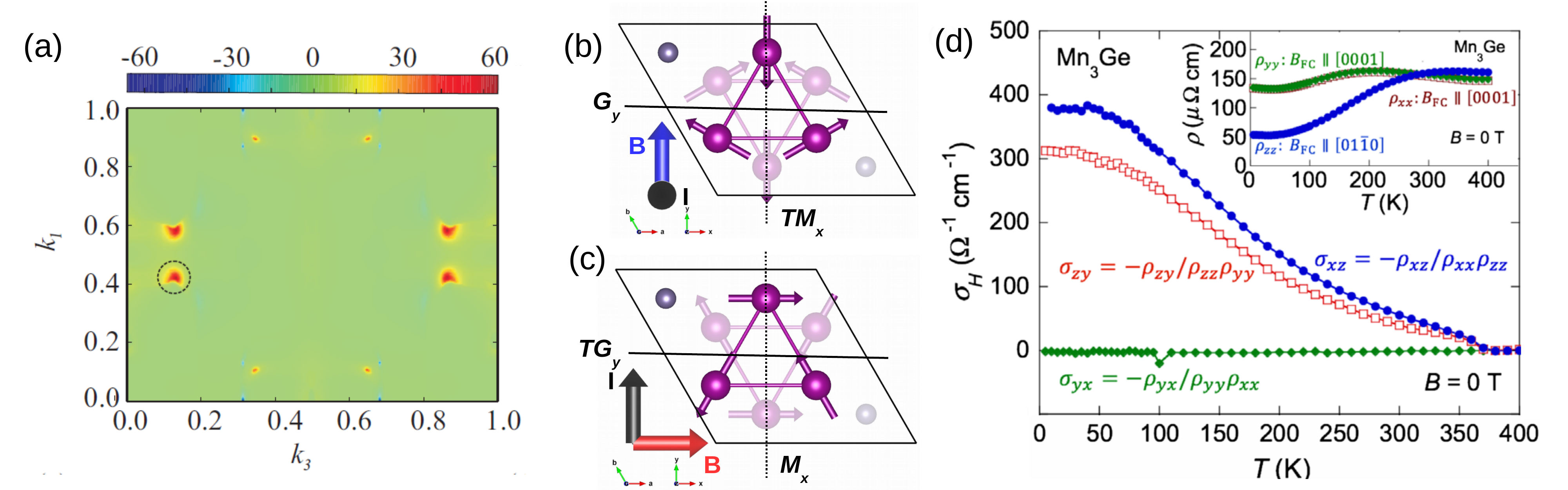}
\caption{\textbf{Anomalous Hall effect in non-collinear antiferromagnets.} \textit{Ab initio} calculation of the Berry curvature reveals the dominating contribution from node lines close to the Weyl points along the $K - M$ axis (see Fig.~\ref{mn3ge}). (b), (c) two different orientation of the   applied electric current-response-external B-field/Berry curvature and its correlation to the non-collinear AF order. (d) Measurement of the temperature dependence of the anomalous Hall effect from the chiral AF spin texture. The colors corresponds to  (b-c). Fig.~ (a) after \cite{Zhang2016d}, and (d)  \cite{Kiyohara2015}.}
\label{mn3ge_exp}
\end{center}
\end{figure*}  

\subsection{Anomalous Hall effect in non-collinear antiferromagnets}
Usually the AHE arises from the presence of magnetization and SOC \cite{Nagaosa2010,Chen2014}. Within this picture one can argue that the AHE is linear in magnetization, as seen from the empirical Eq.~\eqref{hall_exp}, and thus in bipartite AFs should vanish. Indeed, this is true in simple collinear AFs with the combined  time reversal symmetry and half-magnetic unit cell translation $\mathcal{T}T_{1/2}$,  which implies that the Berry curvature is an odd function of crystal momentum when replacing $\mathcal{T}$ with $\mathcal{T}T_{1/2}$ in Eq.~\eqref{Eq_Berry_symmetry} and the AHE vanishes due to the Eq.~\eqref{Eq_Kubo}. We can ask whether it is possible to observe the AHE in systems with a zero net magnetization, or with a zero SOC. The answer to both of these questions is yes \cite{Shindou2001,Bruno2004}. 
 
We start with the AHE in a system with a zero net magnetic moment. In certain AF textures it is not possible to combine the broken $\mathcal{T}$ symmetry with another symmetry operation which would recover the symmetry of the AF and would make the AHE vanish. Indeed, nonzero AHE was predicted  \cite{Haldane1988,Shindou2001,Chen2014,Kubler2014}  and later experimentally discovered in certain disordered and ordered AFs  \cite{Machida2010,Nakatsuji2015,Nayak2016,Kiyohara2015}.
Recent theoretical works  predicted strong AHE in the non-collinear AFs Mn$_{3}$Ir \cite{Chen2014} and Mn$_{3}$Ge \cite{Kubler2014} based on \textit{ab initio} calculations of the intrinsic part of the AHE.   
The largest contribution to the AHE originates from the avoided crossings near the Fermi surface  \cite{Chen2014,Zhang2016d}. As an example, we can compare the \textbf{k}-resolved Berry curvature in Fig.~\ref{mn3ge_exp}(a) with the band structure along the $K-M$ axis  \cite{Kubler2014,Zhang2016d} in the BZ in  Fig.~\ref{mn3ge}(b) for Mn$_{3}$Ge. The detailed interplay of Weyl points and the AHE and the SHE in the Mn$_{3}$X family of AFs was also recently investigated \cite{Yang2017,Zhang2016d}.  

Since the Berry curvature is an axial vector (see Eq.~\eqref{Eq_Berry}), it transforms in the same way as a magnetic moment under the (effective) time-reversal symmetry (see  Eq.~\eqref{Eq_Berry_symmetry}).
Indeed, the magnetic space groups of the candidates Mn$_{3}$Ir and Mn$_{3}$Ge allow for a nonzero magnetic moment, which however does not contribute significantly to the AHE \cite{Chen2014,Kiyohara2015}. The small magnetization allowed for the observation of the AHE  in Mn$_{3}$Ge by orienting the magnetic domains by magnetic fields \cite{Nakatsuji2015,Nayak2016}. The giant magnetic anisotropy energy in Mn$_{3}$Ir has prevented the observation of the AHE in this material untill now  \cite{Zhang2016f}.
Hence, we focus here on the Weyl metal candidate Mn$_{3}$Ge.

The chiral magnetic structure depends on the external magnetic field $\textbf{B}_{\text{FC}}$ applied during sample cooling. Independently on the orientation of the chiral magnetic order, Mn$_{3}$Ge has effective the time reversal symmetry $ \mathcal{T}\mathcal{M}_{z}$, where $\mathcal{M}_{z}$ is the mirror (001) plane symmetry, which implies
$\sigma_{xy}=0$  ($\mathcal{T}\rightarrow  \mathcal{T}\mathcal{M}_{z}$ in  Eq.~\eqref{Eq_Berry_symmetry}). Similarly, for the chiral structure in Fig.~\ref{mn3ge_exp}(b) stabilized by $\textbf{B}_{\text{FC}}\parallel$[010] \cite{Tomiyoshi1982,Kiyohara2015} there is an effective time reversal symmetry $\mathcal{T}\mathcal{M}_{x}$, implying $\sigma_{yz}=0$ and only $\sigma_{xz}\neq 0$ (blue line in Fig.~\ref{mn3ge_exp}(d)). Finally, for the magnetic order in Fig.~\ref{mn3ge_exp}~(c) induced by $\textbf{B}_{\text{FC}}\parallel$[100] \cite{Tomiyoshi1982,Kiyohara2015}, the effective $\mathcal{T}\mathcal{G}_{y}$ symmetry gives $\sigma_{xz}=0$ and only $\sigma_{zy}\neq 0$ (red line in Fig.~\ref{mn3ge_exp}(d)). 
This explains the recent experimental findings summarized in  Fig.~\ref{mn3ge_exp}(d). Kiyohara et al. \cite{Kiyohara2015} reported a nonzero contribution to the AHE from the chiral AF texture, Fig.~\ref{mn3ge_exp}(b), $\sigma^{AF}_{xz}$. For the spin structure in Fig.~\ref{mn3ge_exp}(c) the authors measured a nonzero response only for $\sigma^{AF}_{zy}$. 
In their measurements the anomalous Hall resistivity was extracted from  Eq.~\eqref{hall_exp} by including the contribution from the AF texture $\rho_{H}^{AF}$:
\begin{equation}
\rho_{H}=R_{0}H_{z}+R_{S}M+\rho_{H}^{AF},
\label{hall_exp_af}
\end{equation} 
where $M$ is the net magnetization. \textit{Ab initio} calculations for  Mn$_{3}$Ge give $\sigma_{xz}^{AF}\approx$ 330\,$\Omega^{-1}\text{cm}^{-1}$ \cite{Yang2017}, while the experimentally inferred value is, $\sigma_{xz}$ $\approx$380\,$\Omega^{-1}\text{cm}^{-1}$ \cite{Kiyohara2015}. A strong AHE was also observed in  the GdPtBi AF at very low temperatures, where the authors attributed the effect to the Berry curvature induced by the external magnetic field canting of the AF sublattices  \cite{Suzuki2016}.
 
The discovery of the AHE in non-collinear AFs illustrates how basic research in AFs can advance the general understanding of  spin-dependent transport effects. 
As mentioned above, the AHE without  SOC was also observed in AFs. Several works have identified the so called topological Hall effect originating from  nontrivial magnetic textures where the role of the SOC is overtaken by the spin-chirality  \cite{Surgers2014,Surgers2016}.

\begin{figure*}[t]
\begin{tabular}{|c|c|c|c|c|}
\hline
Antiferromagnet & Phase & T$_{\text{N}}$ & Space group  & Representative effect \\ \hline
\textbf{CuMnAs} & semimetal & 480 \cite{Wadley2015a} & P4/nmm & NSOT control of magnetization \cite{Wadley2016} \\ \hline
\textbf{CuMnAs } & \textit{Dirac semimetal}\cite{Maca2012,Tang2016,Smejkal2016} & $\sim$400 \cite{Maca2012}  & Pnma &  \textit{TopoMIT, TopoAMR} \cite{Smejkal2016} \\ \hline
\textbf{Mn$_{\text{3}}$Ir} & \textit{Weyl metal} & 1000\cite{Kohn2013} & Pm$\overline{3}$m & \textit{AHE}\cite{Chen2014} \\ \hline
\textbf{Mn$_{\text{3}}$Ge}  & Weyl (semi)metal \cite{Yang2017} & 380 \cite{Nayak2016} & P63/mmc & AHE controlled by magnetic field\cite{Kiyohara2015,Nayak2016} \\ \hline
\textbf{SrMnBi$_{\text{2}}$} & Dirac metal\cite{Park2011a} & 290 \cite{Park2011a} & I4/mmm & Angular dependent magnetoresistance \cite{Wang2011e} \\ \hline
\textbf{CaMnBi$_{\text{2}}$} & Dirac metal & 300 \cite{Guo2014} & P4/nmm &  Dirac fermions coupled to magnetism \cite{Guo2014} \\ \hline
EuMnBi$_{\text{2}}$ & Dirac metal & 22* \cite{Masuda2016} & I4/mmm & QHE controlled by magnetism \cite{Masuda2016} \\ \hline 
GdPtBi & \textit{TI} \cite{Mong2010}/Weyl \cite{Hirschberger2016} & 9 \cite{Suzuki2016} & F$\overline{4}$3m & Large thermopower, AHE \cite{Suzuki2016}, \textit{TI} \cite{Mong2010,Li2011,Li2015} \\ \hline
FeSe & superconductor &  & P4/nmm & QSHE \cite{Wang2016e} \\ \hline
X$_{\text{2}}$ Ir$_{\text{2}}$ O$_{\text{7}}$  & \textit{Weyl semimetal}\cite{Wan2011} &  &  Fd$\overline{3}$m & TopoMIT \cite{Tian2015}, wealth of topo. phases \cite{Wan2011,Kondo2015} \\ \hline
\end{tabular}  
    \caption{List of Dirac and Weyl quasiparticle AF candidates with N\'{e}el temperature, proposed phase, crystal symmetry, and representative effect. Italic fonts signal prediction, while normal fonts experimental evidence. *Eu AF sublattice. 
    }
    \label{table}
\end{figure*}

 \subsection{Topological metal insulator transition and anisotropic magnetoresistance}
Predictions of topological quantum phases in oxide iridates have stimulated research on the TopoMIT \cite{Wan2011,Yang2010a,Tian2015,Kondo2015,Zhang2017}. The TopoMIT can be controlled by large pressures  \cite{Yang2010a}, external magnetic fields  \cite{Tian2015}, strain  \cite{Kondo2015}, or doping in X$_{\text{2}} $Ir$_{\text{2}} $O$_{\text{7}} $ \cite{Wan2011,Zhang2017}. 

Recently, a new concept was theoretically predicted in orthorhombic CuMnAs AF \cite{Smejkal2016}. Here the topoMIT is controlled by the interplay between the N\'{e}el vector and the symmetry protection of the Dirac points. The mechanism is related directly to the relativistic spin control and thus is very different from the mechanisms predicted for the X$_{\text{2}} $Ir$_{\text{2}} $O$_{\text{7}}$ family. Remarkably, in the nonsymmorphic $\mathcal{PT}$ AFs, the NSOT can be used to control the N\'{e}el vector direction and in turn the TopoMIT, as we demonstrated on the simple model in Sec 2.2. We illustrate the staggered symmetry of the NSOT by the blue arrows in  Fig.~\ref{cumnas}(a) for the applied current along the [100] direction. The various phases predicted for the CuMnAs AF and depicted in  Fig.~\ref{cumnas}(c) include an AF topological Dirac semimetal for the N\'{e}el vector  $\textbf{n}\parallel$[001] (the 3D electronic dispersion is plotted in the figure of the abstract), an AF semiconductor for $\textbf{n}\parallel$[101], and an AF Dirac semimetal for $\textbf{n}\parallel$[100] with the Dirac point  along the $\Gamma - X$ line and with a small band gap of approximately $ 1\,\text{meV}$ \cite{Smejkal2016}. 

The possibility of controlling the TopoMIT by the NSOT can lead to novel concepts of spin-dependent transport. 
For instance, the non-equilibrium counterpart of the TopoMIT is the topological anisotropic magnetoresistance (TopoAMR)  \cite{Smejkal2016}. The origin of this effect is in the changes of the Fermi surface topology induced by the reorientation of the AF moments, and the corresponding changes of the magnetic symmetry.

\section{Perspectives and conclusion}
The Dirac/Weyl AFs with appealing properties for  spintronics are summarized in Fig.~\ref{table}. While topological Weyl and Dirac (semi)metals (Mn$_{3}$Ge and CuMnAs respectively) are already extensively explored, a myriad of other topological AF systems have a large potential for future research. For instance, GdPtBi was predicted to be an AF TI \cite{Mong2010,Li2015}, and it was recently reported to host a magnetically induced Weyl semimetal state \cite{Hirschberger2016}. Remarkably, several of the guiding AF symmetries important for spintronics are also shared  with the magnetic structures proposed for several of the Fe-based superconductors, making a link to topological superconductivity, which is beyond the scope of this review \cite{Wang2015g,Xu2016}.  N\'{e}el AF order in a mono-layer of FeSe on SrTiO$_{3}$ was reported to exhibit the AF QSHE \cite{Wang2015g,Wang2016e}.

Many of the topological semimetal effects have  come into focus only very recently and are not fully understood. For example, the role of relativistic quasiparticles in the linear magnetoresistance is controversial  \cite{Khouri2016}, as well as the topological nature of the Fermi arcs in Dirac semimetals  \cite{Kargarian2016}, and the nature of the band crossings \cite{Neupane2013,Akrap2016}. Experimental studies of topological AFs might lead to a deeper understanding of these effects, similarly as the research in non-collinear AFs helped our general understanding of transport effects, such as the AHE.
The exciting challenges for theoretical and computational physics are provided by the fact that the topological AFs live very often at the intersection of different physical regimes, e.g., of strong relativistic and electronic correlation regimes \cite{Wan2011,Shinaoka2015,Zhang2017}.
In conclusion, we have described how  the unique symmetries of AFs allow for combining seemingly incompatible effects on explicit examples that included the NSOT and Dirac quasiparticles, or the AHE in AFs. This opens novel research directions in  topological antiferromagnetic spin-orbitronics.  

We acknowledge support from the Grant Agency of the Charles University no. 280815 and of the Czech Republic no. 14-37427G, the Alexander von Humboldt Foundation, EU ERC Synergy Grant No. 610115, and the Transregional Collaborative Research Center (SFB/TRR) 173 SPIN+X. Access to computing and storage facilities owned by parties and projects contributing to the National Grid Infrastructure MetaCentrum provided under the program "Projects of Large Research, Development, and Innovations Infrastructures" (CESNET LM2015042), is greatly appreciated.

\providecommand{\WileyBibTextsc}{}
\let\textsc\WileyBibTextsc
\providecommand{\othercit}{}
\providecommand{\jr}[1]{#1}
\providecommand{\etal}{~et~al.}


\begin{thebibliography}{[100]}

\bibitem{Sinova2015}
 \textsc{J.~Sinova},  \textsc{S.\,O. Valenzuela},  \textsc{J.~Wunderlich},
  \textsc{C.\,H. Back},  and  \textsc{T.~Jungwirth},
 \jr{Rev. Mod. Phys.} \textbf{87}(4), 1213--1260 (2015).


\bibitem{Novoselov2004}
 \textsc{K.\,S. Novoselov},  \textsc{A.\,K. Geim},  \textsc{S.\,V. Morozov},
  \textsc{D.~Jiang},  \textsc{Y.~Zhang},  \textsc{S.\,V. Dubonos},
  \textsc{I.\,V. Grigorieva},  and  \textsc{A.\,A. Firsov},
 \jr{Science} \textbf{306}, 666--669 (2004).


\bibitem{Kane2005}
 \textsc{C.\,L. Kane} and  \textsc{E.\,J. Mele},
 \jr{Phys. Rev. Lett.} \textbf{95}(22), 226801 (2005).


\bibitem{Moore2010a}
 \textsc{J.\,E. Moore},
 \jr{Nature} \textbf{464}(7286), 194--8 (2010).


\bibitem{Kato2004d}
 \textsc{Y.\,K. Kato},  \textsc{R.\,C. Myers},  \textsc{A.\,C. Gossard},  and
  \textsc{D.\,D. Awschalom},
 \jr{Science} \textbf{306}(dec), 1910--1913 (2004).


\bibitem{Wunderlich2004}
 \textsc{J.~Wunderlich},  \textsc{B.~Kastner},  \textsc{J.~Sinova},  and
  \textsc{T.~Jungwirth},
 \jr{Cond-Mat} (2004).


\bibitem{Chernyshov2009}
 \textsc{A.~Chernyshov},  \textsc{M.~Overby},  \textsc{X.~Liu},  \textsc{J.\,K.
  Furdyna},  \textsc{Y.~Lyanda-Geller},  and  \textsc{L.\,P. Rokhinson},
 \jr{Nat. Phys.} \textbf{5}(9), 656--659 (2009).


\bibitem{Miron2011b}
 \textsc{I.\,M. Miron},  \textsc{K.~Garello},  \textsc{G.~Gaudin},
  \textsc{P.\,J. Zermatten},  \textsc{M.\,V. Costache},  \textsc{S.~Auffret},
  \textsc{S.~Bandiera},  \textsc{B.~Rodmacq},  \textsc{A.~Schuhl},  and
  \textsc{P.~Gambardella},
 \jr{Nature} \textbf{476}(7359), 189--193 (2011).


\bibitem{Liu2012}
 \textsc{L.~Liu},  \textsc{C.\,F. Pai},  \textsc{Y.~Li},  \textsc{H.\,W.
  Tseng},  \textsc{D.\,C. Ralph},  and  \textsc{R.\,A. Buhrman},
 \jr{Science} \textbf{336}(6081), 555--558 (2012).


\bibitem{Ciccarelli2016}
 \textsc{C.~Ciccarelli},  \textsc{L.~Anderson},  \textsc{V.~Tshitoyan},
  \textsc{A.\,J. Ferguson},  \textsc{F.~Gerhard},  \textsc{C.~Gould},
  \textsc{L.\,W. Molenkamp},  \textsc{J.~Gayles},
  \textsc{J.~{\v{Z}}elezn{\'{y}}},  \textsc{L.~{\v{S}}mejkal},
  \textsc{Z.~Yuan},  \textsc{J.~Sinova},  \textsc{F.~Freimuth},  and
  \textsc{T.~Jungwirth},
 \jr{Nat. Phys.} \textbf{12}, 855--861 (2016).


\bibitem{Hesjedal2016}
 \textsc{T.~Hesjedal} and  \textsc{Y.~Chen},
 \jr{Nat. Mater.} \textbf{16}(1), 3--4 (2016).


\bibitem{Fert2013}
 \textsc{A.~Fert},  \textsc{V.~Cros},  and  \textsc{J.~Sampaio},
 \jr{Nat. Nanotechnol.} \textbf{8}(3), 152--156 (2013).


\bibitem{Fan2016b}
 \textsc{Y.~Fan} and  \textsc{K.\,L. Wang},
 \jr{SPIN} \textbf{06}(02), 1640001 (2016).


\bibitem{Masuda2016}
 \textsc{H.~Masuda},  \textsc{H.~Sakai},  \textsc{M.~Tokunaga},
  \textsc{Y.~Yamasaki},  \textsc{A.~Miyake},  \textsc{J.~Shiogai},
  \textsc{S.~Nakamura},  \textsc{S.~Awaji},  \textsc{A.~Tsukazaki},
  \textsc{H.~Nakao},  \textsc{Y.~Murakami},  \textsc{T.\,h. Arima},
  \textsc{Y.~Tokura},  and  \textsc{S.~Ishiwata},
 \jr{Sci. Adv.} \textbf{2}(1), e1501117 (2016).


\bibitem{Fan2014a}
 \textsc{Y.~Fan},  \textsc{P.~Upadhyaya},  \textsc{X.~Kou},  \textsc{M.~Lang},
  \textsc{S.~Takei},  \textsc{Z.~Wang},  \textsc{J.~Tang},  \textsc{L.~He},
  \textsc{L.\,T. Chang},  \textsc{M.~Montazeri},  \textsc{G.~Yu},
  \textsc{W.~Jiang},  \textsc{T.~Nie},  \textsc{R.\,N. Schwartz},
  \textsc{Y.~Tserkovnyak},  and  \textsc{K.\,L. Wang},
 \jr{Nat. Mater.} \textbf{13}(7), 699--704 (2014).


\bibitem{Manchon2014b}
 \textsc{A.\,R. Mellnik},  \textsc{J.\,S. Lee},  \textsc{A.~Richardella},
  \textsc{J.\,L. Grab},  \textsc{P.\,J. Mintun},  \textsc{M.\,H. Fischer},
  \textsc{A.~Vaezi},  \textsc{A.~Manchon},  \textsc{E.\,A. Kim},
  \textsc{N.~Samarth},  and  \textsc{D.\,C. Ralph},
 \jr{Nature} \textbf{511}(7510), 449--451 (2014).


\bibitem{Fan2016}
 \textsc{Y.~Fan},  \textsc{K.\,L. Wang},  \textsc{X.~Kou},
  \textsc{P.~Upadhyaya},  \textsc{Q.~Shao},  \textsc{L.~Pan},
  \textsc{M.~Lang},  \textsc{X.~Che},  \textsc{J.~Tang},
  \textsc{M.~Montazeri},  \textsc{K.~Murata},  \textsc{L.\,T. Chang},
  \textsc{M.~Akyol},  \textsc{G.~Yu},  \textsc{T.~Nie},  \textsc{K.\,L. Wong},
  \textsc{J.~Liu},  \textsc{Y.~Wang},  \textsc{Y.~Tserkovnyak},  and
  \textsc{K.\,L. Wang},
 \jr{Spin} \textbf{11}(11), 352 (2016).


\bibitem{Wadley2016}
 \textsc{P.~Wadley},  \textsc{B.~Howells},  \textsc{J.~Zelezny},
  \textsc{C.~Andrews},  \textsc{V.~Hills},  \textsc{R.\,P. Campion},
  \textsc{V.~Novak},  \textsc{K.~Olejn{\'{i}}k},  \textsc{F.~Maccherozzi},
  \textsc{S.\,S. Dhesi},  \textsc{S.\,Y. Martin},  \textsc{T.~Wagner},
  \textsc{J.~Wunderlich},  \textsc{F.~Freimuth},  \textsc{Y.~Mokrousov},
  \textsc{J.~Kunes},  \textsc{J.\,S. Chauhan},  \textsc{M.\,J. Grzybowski},
  \textsc{A.\,W. Rushforth},  \textsc{K.\,W. Edmonds},  \textsc{B.\,L.
  Gallagher},  and  \textsc{T.~Jungwirth},
 \jr{Science} \textbf{351}, 587--590 (2016).


\bibitem{Smejkal2016}
 \textsc{L.~{\v{S}}mejkal},  \textsc{J.~{\v{Z}}elezn{\'{y}}},
  \textsc{J.~Sinova},  and  \textsc{T.~Jungwirth},
 \jr{arXiv:1610.08107} (2016).


\bibitem{Yang2017}
 \textsc{H.~Yang},  \textsc{Y.~Sun},  \textsc{Y.~Zhang},  \textsc{W.\,J. Shi},
  \textsc{S.\,S.\,P. Parkin},  and  \textsc{B.~Yan},
 \jr{New J. Phys.} \textbf{19}(1), 015008 (2017).


\bibitem{Jungwirth2016}
 \textsc{T.~Jungwirth},  \textsc{X.~Marti},  \textsc{P.~Wadley},  and
  \textsc{J.~Wunderlich},
 \jr{Nat. Nanotechnol.} \textbf{11}(3), 231--241 (2016).


\bibitem{Hasan2010}
 \textsc{M.\,Z. Hasan} and  \textsc{C.~Kane},
 \jr{Rev. Mod. Phys.} \textbf{82}(4), 3045--3067 (2010).


\bibitem{Kosterlitz1972}
 \textsc{J.\,M. Kosterlitz} and  \textsc{D.\,J. Thouless},
 \jr{J. Phys. C Solid State Phys.} \textbf{5}(11), L124--L126 (1972).


\bibitem{Kosterlitz1973}
 \textsc{J.\,M. Kosterlitz} and  \textsc{D.\,J. Thouless},
 \jr{J. Phys. C Solid State Phys.} \textbf{6}(7), 1181--1203 (1973).


\bibitem{Klitzing1980}
 \textsc{K.\,V. Klitzing},  \textsc{G.~Dorda},  and  \textsc{M.~Pepper},
 \jr{Phys. Rev. Lett.} \textbf{45}(6), 494--497 (1980).


\bibitem{Thouless1982}
 \textsc{D.\,J. Thouless},  \textsc{M.~Kohmoto},  \textsc{M.\,P. Nightingale},
  and  \textsc{M.~den Nijs},
 \jr{Phys. Rev. Lett.} \textbf{49}(6), 405--408 (1982).


\bibitem{Thouless1983}
 \textsc{D.\,J. Thouless},
 \jr{Phys. Rev. B} \textbf{27}(10), 6083--6087 (1983).


\bibitem{Nagaosa2010}
 \textsc{N.~Nagaosa},  \textsc{J.~Sinova},  \textsc{S.~Onoda},  \textsc{A.\,H.
  MacDonald},  and  \textsc{N.\,P. Ong},
 \jr{Rev. Mod. Phys.} \textbf{82}(2), 1539--1592 (2010).


\bibitem{Haldane1988}
 \textsc{F.\,D.\,M. Haldane},
 \jr{Phys. Rev. Lett.} \textbf{61}(18), 2015 (1988).


\bibitem{Hsieh2008}
 \textsc{D.~Hsieh},  \textsc{D.~Qian},  \textsc{L.~Wray},  \textsc{Y.~Xia},
  \textsc{Y.\,S. Hor},  \textsc{R.\,J. Cava},  and  \textsc{M.\,Z.
  Hasan},
 \jr{Nature} \textbf{452}(7190), 970--4 (2008).


\bibitem{Zhang2009a}
 \textsc{H.~Zhang},  \textsc{C.\,X. Liu},  \textsc{X.\,L. Qi},
  \textsc{X.~Dai},  \textsc{Z.~Fang},  and  \textsc{S.\,C. Zhang},
 \jr{Nat. Phys.} \textbf{5}(6), 438--442 (2009).


\bibitem{Sun2016a}
 \textsc{Y.~Sun},  \textsc{Y.~Zhang},  \textsc{C.~Felser},  and
  \textsc{B.~Yan},
 \jr{Phys. Rev. Lett.} \textbf{117}(14), 146403 (2016).


\bibitem{Vafek2013}
 \textsc{O.~Vafek} and  \textsc{A.~Vishwanath},
 \jr{Annu. Rev. Condens. Matter Phys.} \textbf{5}(1), 83--112 (2014).


\bibitem{Burkov2016}
 \textsc{A.\,A. Burkov},
 \jr{Nat. Mater.} \textbf{15}(11), 1145--1148 (2016).


\bibitem{Herring1937}
 \textsc{C.~Herring},
 \jr{Phys. Rev.} \textbf{52}(4), 365--373 (1937).


\bibitem{Abrikosov11971}
 \textsc{a.\,a. Abrikosov} and  \textsc{S.\,D. Beneslavskii},
 \jr{Sov. Phys. JETP} \textbf{32}(4), 699 (11971).


\bibitem{Neupane2013}
 \textsc{M.~Neupane},  \textsc{S.~Xu},  \textsc{R.~Sankar},
  \textsc{N.~Alidoust},  \textsc{G.~Bian},  \textsc{C.~Liu},
  \textsc{I.~Belopolski},  \textsc{T.\,R. Chang},  \textsc{H.\,T. Jeng},
  \textsc{H.~Lin},  \textsc{A.~Bansil},  \textsc{F.~Chou},  and  \textsc{M.\,Z.
  Hasan},
 \jr{Nat. Commun.} \textbf{5}(001), 1--7 (2014).


\bibitem{Liu2014e}
 \textsc{Z.\,K. Liu},  \textsc{B.~Zhou},  \textsc{Y.~Zhang},  \textsc{Z.\,J.
  Wang},  \textsc{H.\,M. Weng},  \textsc{D.~Prabhakaran},  \textsc{S.~Mo},
  \textsc{Z.\,X. Shen},  \textsc{Z.~Fang},  \textsc{X.~Dai},
  \textsc{Z.~Hussain},  and  \textsc{Y.\,L. Chen},
 \jr{Science (80-. ).} \textbf{343}(February), 864--867 (2014).


\bibitem{Xu2015b}
 \textsc{S.\,Y. Xu},  \textsc{I.~Belopolski},  \textsc{N.~Alidoust},
  \textsc{M.~Neupane},  \textsc{G.~Bian},  \textsc{C.~Zhang},
  \textsc{R.~Sankar},  \textsc{G.~Chang},  \textsc{Z.~Yuan},  \textsc{C.\,C.
  Lee},  \textsc{S.\,M. Huang},  \textsc{H.~Zheng},  \textsc{J.~Ma},
  \textsc{D.\,S. Sanchez},  \textsc{B.~Wang},  \textsc{A.~Bansil},
  \textsc{F.~Chou},  \textsc{P.\,P. Shibayev},  \textsc{H.~Lin},
  \textsc{S.~Jia},  and  \textsc{M.\,Z. Hasan},
 \jr{Science (80-. ).} \textbf{349}(6248), 613--617 (2015).


\bibitem{Lv2015}
 \textsc{B.\,Q. Lv},  \textsc{H.\,M. Weng},  \textsc{B.\,B. Fu},
  \textsc{X.\,P. Wang},  \textsc{H.~Miao},  \textsc{J.~Ma},
  \textsc{P.~Richard},  \textsc{X.\,C. Huang},  \textsc{L.\,X. Zhao},
  \textsc{G.\,F. Chen},  \textsc{Z.~Fang},  \textsc{X.~Dai},  \textsc{T.~Qian},
   and  \textsc{H.~Ding},
 \jr{Phys. Rev. X} \textbf{5}(3), 031013 (2015).


\bibitem{Wang2012f}
 \textsc{Z.~Wang},  \textsc{Y.~Sun},  \textsc{X.\,Q. Chen},
  \textsc{C.~Franchini},  \textsc{G.~Xu},  \textsc{H.~Weng},  \textsc{X.~Dai},
  and  \textsc{Z.~Fang},
 \jr{Phys. Rev. B} \textbf{85}(19), 195320 (2012).


\bibitem{Wang2013g}
 \textsc{Z.~Wang},  \textsc{H.~Weng},  \textsc{Q.~Wu},  \textsc{X.~Dai},  and
  \textsc{Z.~Fang},
 \jr{Phys. Rev. B - Condens. Matter Mater. Phys.} \textbf{88}(12), 1--6 (2013).


\bibitem{Huang2015}
 \textsc{S.\,M. Huang},  \textsc{S.\,Y. Xu},  \textsc{I.~Belopolski},
  \textsc{C.\,C. Lee},  \textsc{G.~Chang},  \textsc{B.~Wang},
  \textsc{N.~Alidoust},  \textsc{G.~Bian},  \textsc{M.~Neupane},
  \textsc{C.~Zhang},  \textsc{S.~Jia},  \textsc{A.~Bansil},  \textsc{H.~Lin},
  and  \textsc{M.\,Z. Hasan},
 \jr{Nat. Commun.} \textbf{6}, 7373 (2015).


\bibitem{Weng2015}
 \textsc{H.~Weng},  \textsc{C.~Fang},  \textsc{Z.~Fang},  \textsc{B.\,A.
  Bernevig},  and  \textsc{X.~Dai},
 \jr{Phys. Rev. X} \textbf{011029}, 1--10 (2015).


\bibitem{Wieder2016}
 \textsc{B.\,J. Wieder},  \textsc{Y.~Kim},  \textsc{A.\,M. Rappe},  and
  \textsc{C.\,L. Kane},
 \jr{Phys. Rev. Lett.} \textbf{116}(18), 1--5 (2016).


\bibitem{Bradlyn2016}
 \textsc{B.~Bradlyn},  \textsc{J.~Cano},  \textsc{Z.~Wang},  \textsc{M.\,G.
  Vergniory},  \textsc{C.~Felser},  \textsc{R.\,J. Cava},  and  \textsc{B.\,A.
  Bernevig},
 \jr{Science (80-. ).} \textbf{353}(6299), aaf5037--7 (2016).


\othercit
\bibitem{bernevig2013topological}
 \textsc{B.~Bernevig} and  \textsc{T.~Hughes},
Topological Insulators and Topological Superconductors (Princeton University
  Press, 2013).


\bibitem{Yang2014a}
 \textsc{B.\,j. Yang} and  \textsc{N.~Nagaosa},
 \jr{Nat. Commun.} \textbf{5}, 4898 (2014).


\bibitem{Young2012}
 \textsc{S.\,M. Young},  \textsc{S.~Zaheer},  \textsc{J.\,C.\,Y. Teo},
  \textsc{C.\,L. Kane},  \textsc{E.\,J. Mele},  and  \textsc{A.\,M.
  Rappe},
 \jr{Phys. Rev. Lett.} \textbf{108}(14), 1--5 (2012).


\bibitem{Fang2015}
 \textsc{C.~Fang},  \textsc{Y.~Chen},  \textsc{H.\,Y. Kee},  and
  \textsc{L.~Fu},
 \jr{Phys. Rev. B} \textbf{92}(8), 1--5 (2015).


\bibitem{Young2015}
 \textsc{S.\,M. Young} and  \textsc{C.\,L. Kane},
 \jr{Phys. Rev. Lett.} \textbf{115}(12), 1--5 (2015).


\bibitem{Schoop2016}
 \textsc{L.\,M. Schoop},  \textsc{M.\,N. Ali},  \textsc{C.~Stra{\ss}er},
  \textsc{A.~Topp},  \textsc{A.~Varykhalov},  \textsc{D.~Marchenko},
  \textsc{V.~Duppel},  \textsc{S.\,S.\,P. Parkin},  \textsc{B.\,V. Lotsch},
  and  \textsc{C.\,R. Ast},
 \jr{Nat. Commun.} \textbf{7}(May), 11696 (2016).


\othercit
\bibitem{bradley2010mathematical}
 \textsc{C.~Bradley} and  \textsc{A.~Cracknell},
The Mathematical Theory of Symmetry in Solids: Representation Theory for Point
  Groups and Space Groups (OUP Oxford, 2010).


\bibitem{Wan2011}
 \textsc{X.~Wan},  \textsc{A.\,M. Turner},  \textsc{A.~Vishwanath},  and
  \textsc{S.\,Y. Savrasov},
 \jr{Phys. Rev. B} \textbf{83}(20), 205101 (2011).


\bibitem{Wang2016c}
 \textsc{Z.~Wang},  \textsc{M.\,G. Vergniory},  \textsc{S.~Kushwaha},
  \textsc{M.~Hirschberger},  \textsc{E.\,V. Chulkov},  \textsc{A.~Ernst},
  \textsc{N.\,P. Ong},  \textsc{R.\,J. Cava},  and  \textsc{B.\,A.
  Bernevig},
 \jr{Phys. Rev. Lett.} \textbf{117}(23), 1--12 (2016).


\bibitem{Zyuzin2012}
 \textsc{A.\,A. Zyuzin},  \textsc{S.~Wu},  and  \textsc{A.\,A. Burkov},
 \jr{Phys. Rev. B} \textbf{85}(16), 165110 (2012).


\bibitem{Chang2016b}
 \textsc{G.~Chang},  \textsc{B.~Singh},  \textsc{S.\,Y. Xu},  \textsc{G.~Bian},
   \textsc{S.\,M. Huang},  \textsc{C.\,H. Hsu},  \textsc{I.~Belopolski},
  \textsc{N.~Alidoust},  \textsc{D.\,S. Sanchez},  \textsc{H.~Zheng},
  \textsc{H.~Lu},  \textsc{X.~Zhang},  \textsc{Y.~Bian},  \textsc{T.\,R.
  Chang},  \textsc{H.\,T. Jeng},  \textsc{A.~Bansil},  \textsc{H.~Hsu},
  \textsc{S.~Jia},  \textsc{T.~Neupert},  \textsc{H.~Lin},  and  \textsc{M.\,Z.
  Hasan},
 \jr{arxiv.org/1604.02124}(apr) (2016).


\bibitem{Nielsen1981}
 \textsc{H.\,B. Nielsen} and  \textsc{M.~Ninomiya},
 \jr{Phys. Lett. B} \textbf{105}(2-3), 219--223 (1981).


\bibitem{Witten2015}
 \textsc{E.~Witten},
 \jr{arxiv.org/1510.07698}(oct) (2015).


\bibitem{Yang2011b}
 \textsc{K.\,Y. Yang},  \textsc{Y.\,M. Lu},  and  \textsc{Y.~Ran},
 \jr{Phys. Rev. B - Condens. Matter Mater. Phys.} \textbf{84}(7), 075129
  (2011).


\bibitem{Chang2016}
 \textsc{G.~Chang},  \textsc{D.\,S. Sanchez},  \textsc{B.\,J. Wieder},
  \textsc{S.\,Y. Xu},  \textsc{F.~Schindler},  \textsc{I.~Belopolski},
  \textsc{S.\,M. Huang},  \textsc{B.~Singh},  \textsc{D.~Wu},
  \textsc{T.~Neupert},  \textsc{T.\,R. Chang},  \textsc{H.~Lin},  and
  \textsc{M.\,Z. Hasan},
 \jr{arxiv.org/1611.07925}(nov), 26 (2016).


\bibitem{Gresch2016}
 \textsc{D.~Gresch},  \textsc{G.~Aut{\`{e}}s},  \textsc{O.\,V. Yazyev},
  \textsc{M.~Troyer},  \textsc{D.~Vanderbilt},  \textsc{B.\,A. Bernevig},  and
  \textsc{A.\,A. Soluyanov},
 \jr{arxiv.org/1610.08983}(oct) (2016).


\bibitem{Xu2015}
 \textsc{S.\,y. Xu},  \textsc{C.~Liu},  \textsc{S.\,K. Kushwaha},
  \textsc{R.~Sankar},  \textsc{J.\,W. Krizan},  \textsc{I.~Belopolski},
  \textsc{M.~Neupane},  \textsc{G.~Bian},  \textsc{N.~Alidoust},
  \textsc{T.\,r. Chang},  \textsc{H.\,t. Jeng},  \textsc{C.\,y. Huang},
  \textsc{W.\,f. Tsai},  \textsc{H.~Lin},  \textsc{P.\,P. Shibayev},
  \textsc{F.\,c. Chou},  \textsc{R.\,J. Cava},  and  \textsc{M.\,Z.
  Hasan},
 \jr{Science (80-. ).} (2015).


\bibitem{Yang2015d}
 \textsc{L.\,X. Yang},  \textsc{Z.\,K. Liu},  \textsc{Y.~Sun},
  \textsc{H.~Peng},  \textsc{H.\,F. Yang},  \textsc{T.~Zhang},
  \textsc{B.~Zhou},  \textsc{Y.~Zhang},  \textsc{Y.\,F. Guo},
  \textsc{M.~Rahn},  \textsc{D.~Prabhakaran},  \textsc{Z.~Hussain},
  \textsc{S.\,K. Mo},  \textsc{C.~Felser},  \textsc{B.~Yan},  and
  \textsc{Y.\,L. Chen},
 \jr{Nat. Phys.} \textbf{11}(9), 728--732 (2015).


\bibitem{Shekhar2015}
 \textsc{C.~Shekhar},  \textsc{A.\,K. Nayak},  \textsc{Y.~Sun},
  \textsc{M.~Schmidt},  \textsc{M.~Nicklas},  \textsc{I.~Leermakers},
  \textsc{U.~Zeitler},  \textsc{Y.~Skourski},  \textsc{J.~Wosnitza},
  \textsc{Z.~Liu},  \textsc{Y.~Chen},  \textsc{W.~Schnelle},
  \textsc{H.~Borrmann},  \textsc{Y.~Grin},  \textsc{C.~Felser},  and
  \textsc{B.~Yan},
 \jr{Nat. Phys.} \textbf{11}(8), 645--649 (2015).


\bibitem{Ali2014}
 \textsc{M.\,N. Ali},  \textsc{J.~Xiong},  \textsc{S.~Flynn},  \textsc{J.~Tao},
   \textsc{Q.\,D. Gibson},  \textsc{L.\,M. Schoop},  \textsc{T.~Liang},
  \textsc{N.~Haldolaarachchige},  \textsc{M.~Hirschberger},  \textsc{N.\,P.
  Ong},  and  \textsc{R.\,J. Cava},
 \jr{Nature} \textbf{514}(7521), 205--8 (2014).


\bibitem{Arnold2015}
 \textsc{F.~Arnold},  \textsc{C.~Shekhar},  \textsc{S.\,C. Wu},
  \textsc{Y.~Sun},  \textsc{R.\,D. dos Reis},  \textsc{N.~Kumar},
  \textsc{M.~Naumann},  \textsc{M.\,O. Ajeesh},  \textsc{M.~Schmidt},
  \textsc{A.\,G. Grushin},  \textsc{J.\,H. Bardarson},  \textsc{M.~Baenitz},
  \textsc{D.~Sokolov},  \textsc{H.~Borrmann},  \textsc{M.~Nicklas},
  \textsc{C.~Felser},  \textsc{E.~Hassinger},  and  \textsc{B.~Yan},
 \jr{Nat. Commun.} \textbf{7}(8), 11615 (2016).


\bibitem{Liang2014}
 \textsc{T.~Liang},  \textsc{Q.~Gibson},  \textsc{M.\,N. Ali},
  \textsc{M.~Liu},  \textsc{R.\,J. Cava},  and  \textsc{N.\,P. Ong},
 \jr{Nat. Mater.} \textbf{14}(3), 280--284 (2014).


\bibitem{Hirschberger2016}
 \textsc{M.~Hirschberger},  \textsc{S.~Kushwaha},  \textsc{Z.~Wang},
  \textsc{Q.~Gibson},  \textsc{S.~Liang},  \textsc{C.\,A. Belvin},
  \textsc{B.\,A. Bernevig},  \textsc{R.\,J. Cava},  and  \textsc{N.\,P.
  Ong},
 \jr{Nat. Mater.} \textbf{15}(11), 1161--1165 (2016).


\bibitem{Jia2016}
 \textsc{S.~Jia},  \textsc{S.\,Y. Xu},  and  \textsc{M.\,Z. Hasan},
 \jr{Nat. Mater.} \textbf{15}(11), 1140--1144 (2016).


\bibitem{Zyuzin2012a}
 \textsc{A.\,A. Zyuzin} and  \textsc{A.\,A. Burkov},
 \jr{Phys. Rev. B} \textbf{86}(11), 115133 (2012).


\bibitem{Jungwirth2014}
 \textsc{T.~Jungwirth},  \textsc{J.~Wunderlich},  \textsc{V.~Nov{\'{a}}k},
  \textsc{K.~Olejn{\'{i}}k},  \textsc{B.\,L. Gallagher},  \textsc{R.\,P.
  Campion},  \textsc{K.\,W. Edmonds},  \textsc{A.\,W. Rushforth},
  \textsc{A.\,J. Ferguson},  and  \textsc{P.~N{\v{e}}mec},
 \jr{Rev. Mod. Phys.} \textbf{86}(3), 855--896 (2014).


\bibitem{Ralph2008}
 \textsc{D.~Ralph} and  \textsc{M.\,D. Stiles},
 \jr{J. Magn. Magn. Mater.} \textbf{320}(7), 1190--1216 (2008).


\bibitem{MacDonald2011}
 \textsc{A.\,H. MacDonald} and  \textsc{M.~Tsoi},
 \jr{Philos. Trans. R. Soc. A Math. Phys. Eng. Sci.} \textbf{369}(1948),
  3098--3114 (2011).


\bibitem{Baltz2016}
 \textsc{V.~Baltz},  \textsc{A.~Manchon},  \textsc{M.~Tsoi},
  \textsc{T.~Moriyama},  \textsc{T.~Ono},  and  \textsc{Y.~Tserkovnyak},
 \jr{arXiv:1606.04284}(jun), 1--138 (2016).


\bibitem{Park2011b}
 \textsc{B.\,G. Park},  \textsc{J.~Wunderlich},  \textsc{X.~Mart{\'{i}}},
  \textsc{V.~Hol{\'{y}}},  \textsc{Y.~Kurosaki},  \textsc{M.~Yamada},
  \textsc{H.~Yamamoto},  \textsc{A.~Nishide},  \textsc{J.~Hayakawa},
  \textsc{H.~Takahashi},  \textsc{A.\,B. Shick},  and
  \textsc{T.~Jungwirth},
 \jr{Nat. Mater.} \textbf{10}(5), 347--51 (2011).


\bibitem{wang2014c}
 \textsc{C.~Wang},  \textsc{H.~Seinige},  \textsc{G.~Cao},  \textsc{J.\,S.
  Zhou},  \textsc{J.\,B. Goodenough},  and  \textsc{M.~Tsoi},
 \jr{Phys. Rev. X} \textbf{4}, 041034 (2014).


\bibitem{Marti2014}
 \textsc{X.~Marti},  \textsc{I.~Fina},  \textsc{C.~Frontera},  \textsc{J.~Liu},
   \textsc{P.~Wadley},  \textsc{Q.~He},  \textsc{R.\,J. Paull},  \textsc{J.\,D.
  Clarkson},  \textsc{J.~Kudrnovsk{\'{y}}},  \textsc{I.~Turek},
  \textsc{J.~Kune{\v{s}}},  \textsc{D.~Yi},  \textsc{J.\,H. Chu},
  \textsc{C.\,T. Nelson},  \textsc{L.~You},  \textsc{E.~Arenholz},
  \textsc{S.~Salahuddin},  \textsc{J.~Fontcuberta},  \textsc{T.~Jungwirth},
  and  \textsc{R.~Ramesh},
 \jr{Nat. Mater.} \textbf{13}(4), 367--374 (2014).


\bibitem{Zelezny2014}
 \textsc{J.~{\v{Z}}elezn{\'{y}}},  \textsc{H.~Gao},
  \textsc{K.~V{\'{y}}born{\'{y}}},  \textsc{J.~Zemen},
  \textsc{J.~Ma{\v{s}}ek},  \textsc{A.~Manchon},  \textsc{J.~Wunderlich},
  \textsc{J.~Sinova},  and  \textsc{T.~Jungwirth},
 \jr{Phys. Rev. Lett.} \textbf{113}(15), 157201 (2014).


\bibitem{He2016}
 \textsc{Q.\,L. He},  \textsc{X.~Kou},  \textsc{A.\,J. Grutter},
  \textsc{G.~Yin},  \textsc{L.~Pan},  \textsc{X.~Che},  \textsc{Y.~Liu},
  \textsc{T.~Nie},  \textsc{B.~Zhang},  \textsc{S.\,M. Disseler},
  \textsc{B.\,J. Kirby},  \textsc{W.~{Ratcliff II}},  \textsc{Q.~Shao},
  \textsc{K.~Murata},  \textsc{X.~Zhu},  \textsc{G.~Yu},  \textsc{Y.~Fan},
  \textsc{M.~Montazeri},  \textsc{X.~Han},  \textsc{J.\,A. Borchers},  and
  \textsc{K.\,L. Wang},
 \jr{Nat. Mater.} \textbf{16}(1), 94--100 (2016).


\bibitem{Ghosh2016}
 \textsc{S.~Ghosh} and  \textsc{A.~Manchon},
 \jr{arxiv.org/1609.01174}(sep) (2016).


\bibitem{Hanke2017a}
 \textsc{J.\,P. Hanke},  \textsc{F.~Freimuth},  \textsc{C.~Niu},
  \textsc{S.~Bl{\"{u}}gel},  and  \textsc{Y.~Mokrousov},
 \jr{arxiv.org/1701.08050}(c) (2017).


\bibitem{Mong2010}
 \textsc{R.\,S.\,K. Mong},  \textsc{A.\,M. Essin},  and  \textsc{J.\,E.
  Moore},
 \jr{Phys. Rev. B} \textbf{81}(24), 1--10 (2010).


\bibitem{Fang2013}
 \textsc{C.~Fang},  \textsc{M.\,J. Gilbert},  and  \textsc{B.\,A.
  Bernevig},
 \jr{Phys. Rev. B} \textbf{88}(8), 085406 (2013).


\bibitem{Tang2016}
 \textsc{P.~Tang},  \textsc{Q.~Zhou},  \textsc{G.~Xu},  and  \textsc{S.\,C.
  Zhang},
 \jr{Nat. Phys.} \textbf{12}, 1100--1104 (2016).


\othercit
\bibitem{Herring1966}
 \textsc{C.~Herring},
{Magnetism: Exchange interactions among itinerant electrons},
 in: Magnetism, edited by G.\,T. Rado and H.~Suhl,  (Academic Press, 1966).


\bibitem{Chen2014}
 \textsc{H.~Chen},  \textsc{Q.~Niu},  and  \textsc{A.\,H. MacDonald},
 \jr{Phys. Rev. Lett.} \textbf{112}(jan), 017205 (2014).


\bibitem{Yang2016b}
 \textsc{B.\,J. Yang},  \textsc{T.\,A. Bojesen},  \textsc{T.~Morimoto},  and
  \textsc{A.~Furusaki},
 \jr{arxiv.org/1604.00843} pp.\,1--32 (2016).


\bibitem{Zelezny2017}
 \textsc{J.~Zelezny},  \textsc{H.~Gao},  \textsc{A.~Manchon},
  \textsc{F.~Freimuth},  \textsc{Y.~Mokrousov},  \textsc{J.~Zemen},
  \textsc{J.~Masek},  \textsc{J.~Sinova},  and  \textsc{T.~Jungwirth},
 \jr{Phys. Rev. B} \textbf{95}, 014403 (2017).


\bibitem{Jones2015}
 \textsc{R.\,O. Jones},
 \jr{Rev. Mod. Phys.} \textbf{87}(3), 897--923 (2015).


\othercit
\bibitem{bluegel2014computing}
 \textsc{S.~Bl{\"u}gel},  \textsc{I.~f{\"u}r Festk{\"o}rperforschung
  (J{\"u}lich). Spring~School},  and  \textsc{I.~for Advanced~Simulation},
Computing Solids: Models, Ab-initio Methods and Supercomputing ; Lecture Notes
  of the 45th IFF Spring School 2014, Lectures notes of the ... IFF Spring
  School (Forschungszentrum, Zentralbibliothek, 2014).


\bibitem{Batabyal2016}
 \textsc{R.~Batabyal},  \textsc{N.~Morali},  \textsc{N.~Avraham},
  \textsc{Y.~Sun},  \textsc{M.~Schmidt},  \textsc{C.~Felser},
  \textsc{A.~Stern},  \textsc{B.~Yan},  and  \textsc{H.~Beidenkopf},
 \jr{Sci. Adv.} \textbf{2}(8), e1600709--e1600709 (2016).


\bibitem{Gradhand2012}
 \textsc{M.~Gradhand},  \textsc{D.\,V. Fedorov},  \textsc{F.~Pientka},
  \textsc{P.~Zahn},  \textsc{I.~Mertig},  and  \textsc{B.\,L.
  Gy{\"{o}}rffy},
 \jr{J. Phys. Condens. Matter} \textbf{24}, 213202 (2012).


\bibitem{Freimuth2015}
 \textsc{F.~Freimuth},  \textsc{S.~Bl{\"{u}}gel},  and
  \textsc{Y.~Mokrousov},
 \jr{Phys. Rev. B} \textbf{92}(6), 064415 (2015).


\bibitem{Liu2011c}
 \textsc{Y.~Liu},  \textsc{A.\,a. Starikov},  \textsc{Z.~Yuan},  and
  \textsc{P.\,J. Kelly},
 \jr{Phys. Rev. B} \textbf{84}(1), 014412 (2011).


\bibitem{Kodderitzsch2013}
 \textsc{D.~K{\"{o}}dderitzsch},  \textsc{K.~Chadova},
  \textsc{J.~Min{\'{a}}r},  and  \textsc{H.~Ebert},
 \jr{New J. Phys.} \textbf{15}(002) (2013).


\bibitem{Janson2014}
 \textsc{O.~Janson},  \textsc{I.~Rousochatzakis},  \textsc{A.\,A. Tsirlin},
  \textsc{M.~Belesi},  \textsc{A.\,A. Leonov},  \textsc{U.\,K.
  R{\"{o}}{\ss}ler},  \textsc{J.~van\,den Brink},  and
  \textsc{H.~Rosner},
 \jr{Nat. Commun.} \textbf{5}(1), 5376 (2014).


\bibitem{Li2017}
 \textsc{G.~Li},  \textsc{B.~Yan},  \textsc{Z.~Wang},  and
  \textsc{K.~Held},
 \jr{Phys. Rev. B} \textbf{95}(3), 035102 (2017).


\bibitem{Perdew1996}
 \textsc{J.\,P. Perdew},  \textsc{K.~Burke},  and
  \textsc{M.~Ernzerhof},
 \jr{Phys. Rev. Lett.} \textbf{77}(18), 3865--3868 (1996).


\bibitem{Park2011}
 \textsc{S.\,R. Park},  \textsc{C.\,H. Kim},  \textsc{J.~Yu},  \textsc{J.\,H.
  Han},  and  \textsc{C.~Kim},
 \jr{Phys. Rev. Lett.} \textbf{107}(15), 156803 (2011).


\bibitem{Lee2013b}
 \textsc{G.~Lee},  \textsc{M.\,A. Farhan},  \textsc{J.\,S. Kim},  and
  \textsc{J.\,H. Shim},
 \jr{Phys. Rev. B} \textbf{87}(24), 245104 (2013).


\bibitem{Park2011a}
 \textsc{J.~Park},  \textsc{G.~Lee},  \textsc{F.~Wolff-Fabris},  \textsc{Y.\,Y.
  Koh},  \textsc{M.\,J. Eom},  \textsc{Y.\,K. Kim},  \textsc{M.\,A. Farhan},
  \textsc{Y.\,J. Jo},  \textsc{C.~Kim},  \textsc{J.\,H. Shim},  and
  \textsc{J.\,S. Kim},
 \jr{Phys. Rev. Lett.} \textbf{107}(12), 1--5 (2011).


\bibitem{Guo2014}
 \textsc{Y.\,F. Guo},  \textsc{A.\,J. Princep},  \textsc{X.~Zhang},
  \textsc{P.~Manuel},  \textsc{D.~Khalyavin},  \textsc{I.\,I. Mazin},
  \textsc{Y.\,G. Shi},  and  \textsc{A.\,T. Boothroyd},
 \jr{Phys. Rev. B} \textbf{90}(7), 075120 (2014).


\bibitem{Wang2011e}
 \textsc{K.~Wang},  \textsc{D.~Graf},  \textsc{H.~Lei},  \textsc{S.\,W. Tozer},
   and  \textsc{C.~Petrovic},
 \jr{Phys. Rev. B - Condens. Matter Mater. Phys.} \textbf{84}(22), 220401(R)
  (2011).


\bibitem{Feng2014}
 \textsc{Y.~Feng},  \textsc{Z.~Wang},  \textsc{C.~Chen},  \textsc{Y.~Shi},
  \textsc{Z.~Xie},  \textsc{H.~Yi},  \textsc{A.~Liang},  \textsc{S.~He},
  \textsc{J.~He},  \textsc{Y.~Peng},  \textsc{X.~Liu},  \textsc{Y.~Liu},
  \textsc{L.~Zhao},  \textsc{G.~Liu},  \textsc{X.~Dong},  \textsc{J.~Zhang},
  \textsc{C.~Chen},  \textsc{Z.~Xu},  \textsc{X.~Dai},  \textsc{Z.~Fang},  and
  \textsc{X.\,J. Zhou},
 \jr{Sci. Rep.} \textbf{4}, 5385 (2014).


\bibitem{Ray2016}
 \textsc{S.\,J. Ray} and  \textsc{L.~Alff},
 \jr{Phys. status solidi} \textbf{254}(1), 1600163 (2017).


\bibitem{Sushkov2015}
 \textsc{A.\,B. Sushkov},  \textsc{J.\,B. Hofmann},  \textsc{G.\,S. Jenkins},
  \textsc{J.~Ishikawa},  \textsc{S.~Nakatsuji},  \textsc{S.~{Das Sarma}},  and
  \textsc{H.\,D. Drew},
 \jr{Phys. Rev. B - Condens. Matter Mater. Phys.} \textbf{92}(24), 241108(R)
  (2015).


\bibitem{Sandratskii1996}
 \textsc{L.\,M. Sandratskii} and  \textsc{J.~K{\"{u}}bler},
 \jr{Phys. Rev. Lett.} \textbf{76}(26), 4963--4966 (1996).


\bibitem{Duan2015}
 \textsc{T.\,F. Duan},  \textsc{W.\,J. Ren},  \textsc{W.\,L. Liu},
  \textsc{S.\,J. Li},  \textsc{W.~Liu},  and  \textsc{Z.\,D. Zhang},
 \jr{Appl. Phys. Lett.} \textbf{107}(8), 82403 (2015).


\bibitem{Tomiyoshi1982}
 \textsc{S.~Tomiyoshi} and  \textsc{Y.~Yamaguchi},
 \jr{J. Phys. Soc. Japan} (1982).


\bibitem{Fujita2016}
 \textsc{H.~Fujita},
 \jr{Phys. status solidi - Rapid Res. Lett.}(dec) (2016).


\bibitem{Maca2012}
 \textsc{F.~M{\'{a}}ca},  \textsc{J.~Ma{\v{s}}ek},  \textsc{O.~Stelmakhovych},
  \textsc{X.~Mart{\'{i}}},  \textsc{H.~Reichlov{\'{a}}},
  \textsc{K.~Uhl{\'{i}}řov{\'{a}}},  \textsc{P.~Beran},  \textsc{P.~Wadley},
  \textsc{V.~Nov{\'{a}}k},  and  \textsc{T.~Jungwirth},
 \jr{J. Magn. Magn. Mater.} \textbf{324}(8), 1606--1612 (2012).


\bibitem{Bernevig2015}
 \textsc{B.\,A. Bernevig},
 \jr{Nat. Phys.} \textbf{11}(9), 698--699 (2015).


\bibitem{Zhang2016g}
 \textsc{A.~Zhang},  \textsc{C.~Liu},  \textsc{C.~Yi},  \textsc{G.~Zhao},
  \textsc{T.\,l. Xia},  \textsc{J.~Ji},  \textsc{Y.~Shi},  \textsc{R.~Yu},
  \textsc{X.~Wang},  \textsc{C.~Chen},  and  \textsc{Q.~Zhang},
 \jr{Nat. Commun.} \textbf{7}, 13833 (2016).


\bibitem{Zhang2016d}
 \textsc{Y.~Zhang},  \textsc{Y.~Sun},  \textsc{H.~Yang},
  \textsc{J.~{\v{Z}}elezn{\'{y}}},  \textsc{S.\,P.\,P. Parkin},
  \textsc{C.~Felser},  and  \textsc{B.~Yan},
 \jr{arXiv:1610.04034} (2016).


\bibitem{Kiyohara2015}
 \textsc{N.~Kiyohara},  \textsc{T.~Tomita},  and  \textsc{S.~Nakatsuji},
 \jr{Phys. Rev. Appl.} \textbf{5}(jun), 064009 (2016).


\bibitem{Shindou2001}
 \textsc{R.~Shindou} and  \textsc{N.~Nagaosa},
 \jr{Phys. Rev. Lett.} \textbf{87}(11), 116801 (2001).


\bibitem{Bruno2004}
 \textsc{P.~Bruno},  \textsc{V.\,K. Dugaev},  and
  \textsc{M.~Taillefumier},
 \jr{Phys. Rev. Lett.} \textbf{93}(9), 096806 (2004).


\bibitem{Kubler2014}
 \textsc{J.~K{\"{u}}bler} and  \textsc{C.~Felser},
 \jr{Europhys. Lett.} \textbf{108}(6), 67001 (2014).


\bibitem{Machida2010}
 \textsc{Y.~Machida},  \textsc{S.~Nakatsuji},  \textsc{S.~Onoda},
  \textsc{T.~Tayama},  and  \textsc{T.~Sakakibara},
 \jr{Nature} \textbf{463}(7278), 210--213 (2010).


\bibitem{Nakatsuji2015}
 \textsc{S.~Nakatsuji},  \textsc{N.~Kiyohara},  and  \textsc{T.~Higo},
 \jr{Nature} \textbf{527}, 212 (2015).


\bibitem{Nayak2016}
 \textsc{A.\,K. Nayak},  \textsc{J.\,E. Fischer},  \textsc{Y.~Sun},
  \textsc{B.~Yan},  \textsc{J.~Karel},  \textsc{A.\,C. Komarek},
  \textsc{C.~Shekhar},  \textsc{N.~Kumar},  \textsc{W.~Schnelle},
  \textsc{J.~K{\"{u}}bler},  \textsc{C.~Felser},  \textsc{S.\,S.\,P. Parkin},
  \textsc{J.~Ku~bler},  \textsc{C.~Felser},  and  \textsc{S.\,S.\,P.
  Parkin},
 \jr{Sci. Adv.} \textbf{2}(4), e1501870--e1501870 (2016).


\bibitem{Zhang2016f}
 \textsc{W.~Zhang},  \textsc{W.~Han},  \textsc{S.\,H. Yang},  \textsc{Y.~Sun},
  \textsc{Y.~Zhang},  \textsc{B.~Yan},  and  \textsc{S.\,S.\,P. Parkin},
 \jr{Sci. Adv.} \textbf{2}, e1600759 (2016).


\bibitem{Suzuki2016}
 \textsc{T.~Suzuki},  \textsc{R.~Chisnell},  \textsc{A.~Devarakonda},
  \textsc{Y.\,T. Liu},  \textsc{W.~Feng},  \textsc{D.~Xiao},  \textsc{J.\,W.
  Lynn},  and  \textsc{J.\,G. Checkelsky},
 \jr{Nat. Phys.} \textbf{12}(July), 1119 (2016).


\bibitem{Surgers2014}
 \textsc{C.~S{\"{u}}rgers},  \textsc{G.~Fischer},  \textsc{P.~Winkel},  and
  \textsc{H.\,V. L{\"{o}}hneysen},
 \jr{Nat. Commun.} \textbf{5}, 3400 (2014).


\bibitem{Surgers2016}
 \textsc{C.~S{\"{u}}rgers},  \textsc{W.~Kittler},  \textsc{T.~Wolf},  and
  \textsc{H.\,v. L{\"{o}}hneysen},
 \jr{AIP Adv.} \textbf{6}(5), 055604 (2016).


\bibitem{Wadley2015a}
 \textsc{P.~Wadley},  \textsc{V.~Hills},  \textsc{M.\,R. Shahedkhah},
  \textsc{K.\,W. Edmonds},  \textsc{R.\,P. Campion},  \textsc{V.~Nov{\'{a}}k},
  \textsc{B.~Ouladdiaf},  \textsc{D.~Khalyavin},  \textsc{S.~Langridge},
  \textsc{V.~Saidl},  \textsc{P.~Nemec},  \textsc{A.\,W. Rushforth},
  \textsc{B.\,L. Gallagher},  \textsc{S.\,S. Dhesi},  \textsc{F.~Maccherozzi},
  \textsc{J.~{\v{Z}}elezn{\'{y}}},  and  \textsc{T.~Jungwirth},
 \jr{Sci. Rep.} \textbf{5}, 17079 (2015).


\bibitem{Kohn2013}
 \textsc{a.~Kohn},  \textsc{A.~Kov{\'{a}}cs},  \textsc{R.~Fan},  \textsc{G.\,J.
  McIntyre},  \textsc{R.\,C.\,C. Ward},  and  \textsc{J.\,P. Goff},
 \jr{Sci. Rep.} \textbf{3}(aug), 2412 (2013).


\bibitem{Li2011}
 \textsc{C.~Li},  \textsc{J.\,S. Lian},  and  \textsc{Q.~Jiang},
 \jr{Phys. Rev. B - Condens. Matter Mater. Phys.} \textbf{83}(23), 1--5 (2011).


\bibitem{Li2015}
 \textsc{Z.~Li},  \textsc{H.~Su},  \textsc{X.~Yang},  and
  \textsc{J.~Zhang},
 \jr{Phys. Rev. B} \textbf{91}(23), 235128 (2015).


\bibitem{Wang2016e}
 \textsc{Z.\,F. Wang},  \textsc{H.~Zhang},  \textsc{D.~Liu},  \textsc{C.~Liu},
  \textsc{C.~Tang},  \textsc{C.~Song},  \textsc{Y.~Zhong},  \textsc{J.~Peng},
  \textsc{F.~Li},  \textsc{C.~Nie},  \textsc{L.~Wang},  \textsc{X.\,J. Zhou},
  \textsc{X.~Ma},  \textsc{Q.\,K. Xue},  and  \textsc{F.~Liu},
 \jr{Nat. Mater.} \textbf{15}(September), 968 (2016).


\bibitem{Tian2015}
 \textsc{Z.~Tian},  \textsc{Y.~Kohama},  \textsc{T.~Tomita},
  \textsc{H.~Ishizuka},  \textsc{T.\,H. Hsieh},  \textsc{J.\,J. Ishikawa},
  \textsc{K.~Kindo},  \textsc{L.~Balents},  and  \textsc{S.~Nakatsuji},
 \jr{Nat. Phys.} \textbf{12}(November), 134 (2015).


\bibitem{Kondo2015}
 \textsc{T.~Kondo},  \textsc{M.~Nakayama},  \textsc{R.~Chen},  \textsc{J.\,J.
  Ishikawa},  \textsc{E.\,G. Moon},  \textsc{T.~Yamamoto},  \textsc{Y.~Ota},
  \textsc{W.~Malaeb},  \textsc{H.~Kanai},  \textsc{Y.~Nakashima},
  \textsc{Y.~Ishida},  \textsc{R.~Yoshida},  \textsc{H.~Yamamoto},
  \textsc{M.~Matsunami},  \textsc{S.~Kimura},  \textsc{N.~Inami},
  \textsc{K.~Ono},  \textsc{H.~Kumigashira},  \textsc{S.~Nakatsuji},
  \textsc{L.~Balents},  and  \textsc{S.~Shin},
 \jr{Nat. Commun.} \textbf{6}, 1--8 (2015).


\bibitem{Yang2010a}
 \textsc{B.\,J. Yang} and  \textsc{Y.\,B. Kim},
 \jr{Phys. Rev. B} \textbf{82}(8), 085111 (2010).


\bibitem{Zhang2017}
 \textsc{H.~Zhang},  \textsc{K.~Haule},  and  \textsc{D.~Vanderbilt},
 \jr{Phys. Rev. Lett.} \textbf{118}(2), 026404 (2017).


\bibitem{Wang2015g}
 \textsc{Q.~Wang},  \textsc{Y.~Shen},  \textsc{B.~Pan},  \textsc{X.~Zhang},
  \textsc{K.~Ikeuchi},  \textsc{K.~Iida},  \textsc{A.\,D. Christianson},
  \textsc{H.\,C. Walker},  \textsc{D.\,T. Adroja},  \textsc{M.~Abdel-Hafiez},
  \textsc{X.~Chen},  \textsc{D.\,A. Chareev},  \textsc{A.\,N. Vasiliev},  and
  \textsc{J.~Zhao},
 \jr{Nat. Commun.} \textbf{7}, 1--15 (2015).


\bibitem{Xu2016}
 \textsc{G.~Xu},  \textsc{B.~Lian},  \textsc{P.~Tang},  \textsc{X.\,L. Qi},
  and  \textsc{S.\,C. Zhang},
 \jr{Phys. Rev. Lett.} \textbf{117}(4), 1--5 (2016).


\bibitem{Khouri2016}
 \textsc{T.~Khouri},  \textsc{U.~Zeitler},  \textsc{C.~Reichl},
  \textsc{W.~Wegscheider},  \textsc{N.\,E. Hussey},  \textsc{S.~Wiedmann},  and
   \textsc{J.\,C. Maan},
 \jr{Phys. Rev. Lett.} \textbf{117}(25), 256601 (2016).


\bibitem{Kargarian2016}
 \textsc{M.~Kargarian},  \textsc{M.~Randeria},  and  \textsc{Y.\,M. Lu},
 \jr{Proc. Natl. Acad. Sci. U. S. A.} \textbf{113}(31), 8648--52 (2016).


\bibitem{Akrap2016}
 \textsc{A.~Akrap},  \textsc{M.~Hakl},  \textsc{S.~Tchoumakov},
  \textsc{I.~Crassee},  \textsc{J.~Kuba},  \textsc{M.\,O. Goerbig},
  \textsc{C.\,C. Homes},  \textsc{O.~Caha},  \textsc{J.~Novak},
  \textsc{F.~Teppe},  \textsc{W.~Desrat},  \textsc{S.~Koohpayeh},
  \textsc{L.~Wu},  \textsc{N.\,P. Armitage},  \textsc{A.~Nateprov},
  \textsc{E.~Arushanov},  \textsc{Q.\,D. Gibson},  \textsc{R.\,J. Cava},
  \textsc{D.~{Van Der Marel}},  \textsc{B.\,A. Piot},  \textsc{C.~Faugeras},
  \textsc{G.~Martinez},  \textsc{M.~Potemski},  and  \textsc{M.~Orlita},
 \jr{Phys. Rev. Lett.} \textbf{117}(13), 3--8 (2016).


\bibitem{Shinaoka2015}
 \textsc{H.~Shinaoka},  \textsc{S.~Hoshino},  \textsc{M.~Troyer},  and
  \textsc{P.~Werner},
 \jr{Phys. Rev. Lett.} \textbf{115}(15), 156401 (2015).


\end{thebibliography}
\end{document}